\documentclass[%
 aip,
 jcp,%
 amsmath,amssymb,
 reprint,%
]{revtex4-1}

\usepackage{amsmath}
\usepackage{graphicx}
\usepackage{xspace}
\usepackage{latexsym}
\usepackage{subfig}
\usepackage{braket}
\usepackage{cleveref}
\usepackage{threeparttable}
\usepackage{enumerate}
\usepackage{setspace}
\usepackage{mhchem}
\usepackage[dvipsnames]{xcolor}

\usepackage{array}
\newcolumntype{L}[1]{>{\raggedright\let\newline\\\arraybackslash\hspace{0pt}}m{#1}}
\newcolumntype{C}[1]{>{\centering\let\newline\\\arraybackslash\hspace{0pt}}m{#1}}
\newcolumntype{R}[1]{>{\raggedleft\let\newline\\\arraybackslash\hspace{0pt}}m{#1}}

\crefname{figure}{Figure}{Figures}
\crefname{table}{Table}{Tables}
\crefname{equation}{Eq.}{Eqs.}
\crefname{section}{Section}{Sections}

\newcommand*{\meh}{\ensuremath{mE_h}\xspace}
\newcommand*{\eh}{\ensuremath{E_h}\xspace}

\newcommand{\h}[2]{h_{{#1}}^{{#2}}}
\renewcommand{\th}[2]{\tilde{h}_{{#1}}^{{#2}}}

\newcommand{\vv}[2]{{v}_{{#1}}^{{#2}}}
\renewcommand{\c}[1]{a^\dagger_{#1}}
\renewcommand{\a}[1]{a^{\ }_{#1}}
\newcommand{\f}[2]{f_{{#1}}^{{#2}}}
\newcommand{\e}[1]{\ensuremath{\varepsilon_{#1}}}
\newcommand{\pdm}[2]{\gamma_{{#1}}^{{#2}}}
\renewcommand{\AA}{\r{A}\xspace}

\newcommand{\tmpsnevpt}{t-MPS-NEVPT2\xspace}
\newcommand{\scmpsnevpt}{sc-MPS-NEVPT2\xspace}
\newcommand{\scnevpt}{sc-DMRG-NEVPT2\xspace}
\newcommand{\dmrgscf}{DMRG-SCF\xspace}
\newcommand{\tddmrg}{td-DMRG\xspace}

\newcommand{\bup}{\ensuremath{1B_u^+}\xspace}
\newcommand{\bum}{\ensuremath{1B_u^-}\xspace}
\newcommand{\ag}[1]{\ensuremath{{#1}A_g^-}\xspace}

\begin{document}

\raggedbottom 

\title{Time-dependent $N$-electron valence perturbation theory with matrix product state reference wavefunctions for large active spaces and basis sets: Applications to the chromium dimer and {\it all-trans} polyenes}

\author{Alexander~Yu.~Sokolov}
\email{alexander.y.sokolov@gmail.com}
\affiliation{Division of Chemistry and Chemical Engineering, California Institute of Technology,
Pasadena, CA 91125, USA}
\author{Sheng~Guo}
\affiliation{Division of Chemistry and Chemical Engineering, California Institute of Technology, Pasadena, CA 91125, USA}
\author{Enrico~Ronca}
\affiliation{Division of Chemistry and Chemical Engineering, California Institute of Technology, Pasadena, CA 91125, USA}
\author{Garnet~Kin-Lic~Chan}
\email{gkc1000@gmail.com}
\affiliation{Division of Chemistry and Chemical Engineering, California Institute of Technology, Pasadena, CA 91125, USA}

\begin{abstract}
In earlier work [J. Chem. Phys. 144, 064102 (2016)], we introduced a time-dependent formulation of the second-order $N$-electron valence perturbation theory (t-NEVPT2) which (i) had a lower computational scaling than the usual internally-contracted perturbation formulation, and (ii) yielded the fully uncontracted NEVPT2 energy. Here, we present a combination of t-NEVPT2 with a matrix product state (MPS) reference wavefunction (\tmpsnevpt) that allows to compute uncontracted dynamic correlation energies for large active spaces and basis sets, using the time-dependent density matrix renormalization group (\tddmrg) algorithm. In addition, we report a low-scaling MPS-based implementation of  strongly-contracted NEVPT2 (\scmpsnevpt) that avoids computation of the four-particle reduced density matrix. We use these new methods to compute the dissociation energy of the chromium dimer and to study the low-lying excited states in {\it all-trans} polyenes (\ce{C4H6} to \ce{C24H26}), incorporating dynamic correlation for reference wavefunctions with up to 24 active electrons and orbitals.
\end{abstract}

\titlepage

\maketitle

\section{Introduction}
\label{sec:intro}

Recent advances in molecular electron correlation methods have made it possible to describe dynamic correlation in weakly correlated systems with several hundreds of atoms by taking advantage of accurate local correlation approximations.\cite{Schutz:2001p661,Werner:2003p8149,Riplinger:2013p034106,Riplinger:2013p134101} Similarly, efficient methods for static correlation\cite{White:1999p4127,Legeza2008,Booth:2009p054106,Kurashige:2009p234114,Marti:2011p6750,Chan:2011p465,Wouters:2014p272} now exist that allow us to simulate systems with a large number of strongly correlated open-shell orbitals.\cite{Kurashige:2013p660,Chalupsky:2014p15977,Wouters:2014p241103,Sharma:2014p927} Yet, despite these efforts, the challenge still remains to efficiently describe  dynamic correlation in the presence of significant static correlation. This is a scenario one faces when treating some complicated chemical systems, such as transition metal
compounds with multiple metals.\cite{Kurashige:2013p660,Sharma:2014p927} Technically, the challenges are the high computational cost of the existing canonical algorithms, algebraic complexity of the underlying equations,\cite{MacLeod:2015p051103} as well as numerical instabilities in the simulations.\cite{Evangelisti:1987p4930,Kowalski:2000p757,Kowalski:2000p052506,Evangelista:2014p054109} 

The standard approach to electron correlation in multi-reference (strongly correlated) systems is to first carry out a high-level description of static correlation in a small subset of near-degenerate (active) orbitals, followed by a lower-level description of dynamic correlation in the remaining large set of core and external orbitals. Here, the main challenge is to combine the high-level and low-level descriptions without sacrificing their accuracy, while retaining a low computational cost. For this purpose, most multi-reference dynamic correlation theories use the internal contraction approximation,\cite{Werner:1988p5803,Angeli:2001p10252,Angeli:2001p297} which represents the complicated active-space wavefunction in terms of simpler quantities, such as the reduced density matrices (RDM). While this approximation has been enormously useful in quantum chemistry,
\cite{Hirao:1992p374,Werner:1988p5803,Andersson:1992p1218,Werner:1996p645,Angeli:2001p10252,Angeli:2001p297,Yanai:2006p194106,Yanai:2007p104107,Kurashige:2011p094104,Saitow:2013p044118,Saitow:2015p5120,Datta:2011p214116,Evangelista:2011p114102,Kohn:2012p176,Guo:2016p1583,Freitag:2017p451}
internally-contracted methods require the computation of high-order (three- and four-particle) RDMs. These become prohibitively expensive for larger active spaces.

We have recently proposed a {time-dependent formulation} of multi-reference perturbation theory\cite{Sokolov:2016p064102} that efficiently describes  dynamic correlation by representing the active-space wavefunction in terms of compact time-dependent quantities (active-space imaginary-time Green's functions). This does not introduce any additional approximations, nor do high-order RDMs appear in the equations.
The resulting time-dependent theory is equivalent to the fully {uncontracted} perturbation theory, but in fact displays a lower computational scaling than the internally-contracted approximations, particularly with respect to the number of active orbitals. 

Our previous work described the implementation of the time-dependent second-order $N$-electron valence perturbation theory (t-NEVPT2)
for complete active-space self-consistent field (CASSCF) reference wavefunctions. Here, we present a new implementation combining t-NEVPT2 with matrix product state (MPS) reference wavefunctions, thus allowing to describe dynamic correlation in multi-reference systems with
much larger active spaces. As we will demonstrate, the resulting \tmpsnevpt approach requires computation of up to two-particle imaginary-time Green's functions that can be evaluated using the time-dependent density matrix renormalization group (\tddmrg) algorithm\cite{White:1999p4127,Kurashige:2009p234114,Marti:2011p6750,Chan:2011p465,Wouters:2014p272} in a highly parallel fashion. In addition, for comparison,
we present a low-scaling MPS-based implementation of the strongly-contracted NEVPT2 (\scmpsnevpt),\cite{Angeli:2001p10252,Angeli:2001p297,Guo:2016p1583} which does not require computation of the four-particle density matrix. To demonstrate the capabilities of these new methods, we apply them to compute the dissociation energy of the chromium dimer and to study the low-lying excited states in the {\it all-trans} polyenes.

This paper is organized as follows. We begin with a brief overview of t-NEVPT2 (\cref{sec:t_nevpt2}) and matrix product state wavefunctions (\cref{sec:dmrg}). We then describe the details of our \tmpsnevpt implementation (\cref{sec:t_mps_nevpt2}) and outline the computational details (\cref{sec:comp_details}). In \cref{sec:results}, we use \tmpsnevpt to compute energies along the \ce{N2} dissociation curve (\cref{sec:numerical_accuracy}), the dissociation energy of the chromium dimer (\cref{sec:cr2}), and the vertical excitation energies in {\it all-trans} polyenes (\cref{sec:polyenes}). In Section \ref{sec:conclusions}, we present the conclusions of our work. The details of the efficient \scmpsnevpt implementation are described in the Appendix B.

\section{Theory}
\label{sec:theory}

\subsection{Time-dependent formulation of $N$-electron valence perturbation theory (t-NEVPT2)}
\label{sec:t_nevpt2}

We begin with a brief overview of multi-reference perturbation theory in its time-dependent form. Our starting point is a zeroth-order electronic wavefunction $\ket{\Psi_0}$ computed in a complete active space (CAS) of molecular orbitals. We require $\ket{\Psi_0}$ to be an eigenfunction of a zeroth-order Hamiltonian $\hat{H}^{(0)}$. Following our previous work,\cite{Sokolov:2016p064102} we assume that $\hat{H}^{(0)}$ is the Dyall Hamiltonian,\cite{Dyall:1995p4909} $\hat{H}^{(0)}=\hat{H}_D$, defined as:
\begin{align}
	\label{eq:H_Dyall}
	\hat{H}_D =& \sum_{i}  \e{i} \c{i}\a{i} + \sum_{a}  \e{a} \c{a}\a{a} + \hat{H}_{act} \ ,
\end{align}
where we partition the spin-orbitals into three sets: (i) {\it core} (doubly-occupied) with indices $i,j,k,l$; (ii) {\it active} with indices $u,v,w,x,y,z$; and (iii) {\it external} (unoccupied) with indices $a,b,c,d$. The orbital energies $\e{i}$ and $\e{a}$ are defined as the core and external eigenvalues of the generalized Fock operator,
\begin{align}
	\label{eq:f_gen}
	\f{p}{q} &= \h{p}{q} + \sum_{rs} \vv{pr}{qs} \pdm{s}{r} \ ,
\end{align}
where $\h{p}{q}$ and $\vv{pq}{rs}$ are the usual one- and antisymmetrized two-electron integrals, and $\pdm{q}{p}=\braket{\Psi_0|\c{p}\a{q}|\Psi_0}$ is the one-particle density matrix of $\ket{\Psi_0}$. The $\hat{H}_{act}$ operator describes the two-electron interaction in the active space:
\begin{align}
	\label{eq:H_act}
	\hat{H}_{act} &= \sum_{xy}(\h{x}{y} + \sum_{i} \vv{xi}{yi}) \c{x} \a{y}
	+ \frac{1}{4} \sum_{xywz} \vv{xy}{zw} \c{x} \c{y} \a{w} \a{z} \ .
\end{align}

Having specified $\ket{\Psi_0}$ and $\hat{H}^{(0)}$, we now consider an expansion of the energy with respect to the perturbation $\lambda \hat{V}  = \lambda(\hat{H}-\hat{H}_D)$. 
Rather than expressing the energy in a time-independent perturbation series, as is commonly done in the Rayleigh-Schr\"odinger perturbation theory, we consider a time-dependent expansion\cite{Sokolov:2016p064102} with respect to the perturbation operator $\hat{V}(\tau) = e^{(\hat{H}_D  - E_D)\tau} \, \hat{V} \, e^{-(\hat{H}_D - E_D)\tau}$. 
The second-order energy contribution can be written as:
\begin{align}
	\label{eq:E_tdpt2}
	E^{(2)} 
        &= -\int_{0}^\infty \! \mathrm{d}\tau \braket{ \Psi_0 | \hat{V}^{\prime\dag}(\tau) \hat{V}^\prime(0) |\Psi_0} \notag \\
        &= -\int_{0}^\infty \! \mathrm{d}\tau \braket{ \Psi_0 | \hat{V}^{\prime\dag} e^{-(\hat{H}_D - E_D)\tau} \hat{V}^\prime |\Psi_0} \ , 
\end{align}
where $\tau$ is imaginary time and $\hat{V}^{\prime}$ denotes the part of $\hat{V}$ that contributes to the first-order wavefunction $\ket{\Psi^{(1)}}$ ($\hat{V}^{\prime} = \hat{Q} \hat{V}$, $\hat{Q} = 1 - \ket{\Psi_0}\bra{\Psi_0}$). \cref{eq:E_tdpt2} is the Laplace transform of the second-order energy expression from the Rayleigh-Schr\"odinger perturbation theory, which yields the exact (uncontracted) energy of the second-order $N$-electron valence perturbation theory (NEVPT2).\cite{Angeli:2001p10252,Angeli:2001p297} However, unlike the standard uncontracted theory, the time-dependent energy expression \eqref{eq:E_tdpt2} does not require the costly representation of the first-order wavefunction in a large space of determinants and can be evaluated very efficiently. Expanding the perturbation operator $\hat{V}$ into classes of excitations, the second-order energy can be expressed as a sum of 8 terms related to the numbers of holes and particles
created in the core and external spaces (see Refs.~\citenum{Angeli:2001p10252} and \citenum{Zgid:2009p194107} for a complete definition):
\begin{align}
  \label{eq:e_contributions}
  E^{(2)} 
  &= E^{[0]} 
  + E^{[+1]} 
  + E^{[-1]} 
  + E^{[+2]} 
  + E^{[-2]} \\ \notag
  &+ E^{[+1']} 
  + E^{[-1']} 
  + E^{[0']}  \ .
\end{align}
The central quantities to compute in \cref{eq:e_contributions} are the one- and two-particle imaginary-time Green's functions in the active space (1- and 2-GF). 
Specifically, the $E^{[-1]}$ and $E^{[+1]}$ terms require computation of 1-GFs (e.g., $G(\tau)=\braket{\Psi_0| \c{x} (\tau) \a{y} |\Psi_0}$), while 2-GFs are necessary to compute the $E^{[-2]}$, $E^{[+2]}$, and $E^{[0']}$ contributions (e.g., $G(\tau)=\braket{\Psi_0| \c{w} (\tau) \c{x} (\tau) \a{y} \a{z} |\Psi_0}$). The $E^{[-1']}$ and $E^{[+1']}$ components formally involve the three-particle active-space Green's function, but can be efficiently evaluated by forming intermediates. Defining
\begin{align}
	\label{eq:h_three_particle}
	\hat{h}^{\dag}_{p} &= \sum_{x} \th{x}{p} \c{x} + \frac{1}{2} \sum_{xyz} \vv{xy}{pz} \c{x} \c{y} \a{z} \ , \\
	\th{p}{q} &= \h{p}{q} + \sum_{i} \vv{pi}{qi} \ , 
\end{align}
the $E^{[-1']}$ and $E^{[+1']}$ contributions can be expressed as expectation values of the single-index operators:
\begin{align}
	\label{eq:e_+1p}
	E^{[+1']} 
	&= -\int_{0}^{\infty}  \sum_{i} e^{\e{i} \tau} \braket{\Psi_0| \hat{h}^{}_{i}  (\tau) \hat{h}^{\dag}_{i} |\Psi_0} \mathrm{d}\tau \ , \\
	\label{eq:e_-1p}
	E^{[-1']}
	&= - \int_{0}^{\infty}  \sum_{a} e^{-\e{a} \tau} \braket{\Psi_0| \hat{h}^{\dag}_{a}  (\tau) \hat{h}^{}_{a} |\Psi_0} \mathrm{d}\tau \ .
\end{align}
The bottleneck of the time-dependent NEVPT2 (t-NEVPT2) algorithm, when $\ket{\Psi_0}$ is expanded in determinants, is the evaluation of the 2-GF that has $\mathcal{O}(N_\tau \times N_{det} \times N_{act}^6)$ computational cost, where $N_{act}$ is the number of active orbitals, $N_{det}$ is the dimension of the CAS Hilbert space, and $N_\tau$ is the number of time steps ($N_\tau$ $\sim$ 10-20). As a result, t-NEVPT2 has a significantly lower scaling with $N_{act}$ than the internally-contracted NEVPT2 approaches,\cite{Angeli:2001p10252,Angeli:2001p297} which require computation of up to the four-particle reduced density matrix (4-RDM) with $\mathcal{O}(N_{det} \times N_{act}^8)$ cost.  However, the computational cost of these two types of NEVPT2 formulations still scales exponentially with $N_{act}$, since the number of determinants in the zeroth-order wavefunction $N_{det}\sim \mathcal{O}(e^{N_{act}})$, and solving for $\ket{\Psi_0}$ becomes a bottleneck for large active spaces. In the following sections we will describe how this limitation can be
ameliorated by expressing $\ket{\Psi_0}$ in the matrix product state (MPS) representation.

  \subsection{Matrix product state (MPS) wavefunctions and the density matrix renormalization group (DMRG) algorithm}
  \label{sec:dmrg}

  In this section, we very briefly introduce matrix product state (MPS) wavefunctions, which will be used extensively in this work. Further
details about MPS algorithms can be found in references such as Refs.~\citenum{Schollwock:2005p259} and \citenum{Chan:2016p014102}.
  The MPS  is a nonlinear wavefunction composed of a product of tensors where each tensor corresponds to one (or more) orbitals in the basis. The most common form of the MPS wavefunction is the one-site MPS  written in the following form:
  \begin{align}
  	\label{eq:one_site_mps}
  	\ket{\Psi} = \sum_{n_1\ldots n_k} \mathbf{A}^{n_1} \mathbf{A}^{n_2} \ldots \mathbf{A}^{n_p} \ldots \mathbf{A}^{n_{k}} \ket{n_1n_2\ldots n_k}\ ,
  \end{align}
  where $\ket{n_1n_2\ldots n_k}$ is a Slater determinant, $\mathbf{A}^{n_p}$ is a tensor that corresponds to only one orbital $p$ with occupancy $n_p\in\{\ket{},\ket{\uparrow},\ket{\downarrow},\ket{\uparrow\downarrow}\}$ (also referred as the site $p$), and the total number of tensors $\mathbf{A}^{n_p}$ equals  the number of orbitals $k$ in the basis (for a CAS wavefunction, $k=N_{act}$). For each $p$ in the range $\{2,\ldots,k-1\}$, $\mathbf{A}^{n_p}$ is a matrix with fixed dimensions $M\times M$. The dimensions of $\mathbf{A}^{n_1}$ and $\mathbf{A}^{n_k}$ are $1\times M$ and $M\times1$, respectively. As a result, for a specified set of occupation numbers $n_1n_2\ldots n_k$, the product of tensors in \cref{eq:one_site_mps} yields a scalar corresponding to the coefficient of the determinant $\ket{n_1n_2\ldots n_k}$ in the expansion of the wavefunction $\ket{\Psi}$. The parameter $M$, referred to as the bond dimension, controls the flexibility of the MPS, which increases as $M$ increases.

The most common algorithm when working with MPS is the sweep algorithm, where linear algebra operations are carried out on a single tensor at a time, sweeping
successively through the orbitals $1 \ldots k$. The density matrix renormalization group (DMRG) is the prototypical version of such an algorithm, and corresponds to a variational energy minimization using
a sweep through the tensors. At a given site in the sweep algorithm, the tensor that is being operated on yields
the linear expansion coefficients of the wavefunction in a many-body renormalized basis. For example, at site $p$, we  define 
the left and right renormalized bases as
\begin{widetext}
\begin{align}
\ket{l_{\alpha_p}} &= \sum_{n_1 \ldots n_{p-1}} [\mathbf{A}^{n_1} \mathbf{A}^{n_2} \ldots \mathbf{A}^{n_{p-1}}]_{\alpha_p} \ket{n_1 n_2 \ldots n_{p-1}} \ , \\
\ket{r_{\alpha_{p+1}}} &= \sum_{n_{p+1} \ldots n_k} [\mathbf{A}^{n_{p+1}} \mathbf{A}^{n_{p+2}} \ldots \mathbf{A}^{n_{k}}]_{\alpha_{p+1}} \ket{n_{p+1} n_{p+2} \ldots n_{k} }
\end{align}
\end{widetext}
and the total wavefunction becomes
\begin{align}
\ket{\Psi} = \sum_{l_{\alpha_p} n_p r_{\alpha_{p+1}}} [\mathbf{A}^{n_p}]_{\alpha_p \alpha_{p+1}} \ket{l_{\alpha_p} n_p r_{\alpha_{p+1}}} \ .
\end{align}
In this context, we can think of $\mathbf{A}^{n_p}$ as representing a ``site wavefunction'', and operators projected
into the renormalized space  $ \{ \ket{l_{\alpha_p} n_p r_{\alpha_{p+1}}} \}$ act on this site wavefunction as ``site operators''. 
General numerical algorithms for wavefunctions can be converted into sweep algorithms for MPS by identifying wavefunctions with site wavefunctions, and operators with site operators.
This idea is used in the time-dependent DMRG (\tddmrg) algorithm that we employ below.

\subsection{t-NEVPT2 with MPS reference wavefunctions (\tmpsnevpt)}
\label{sec:t_mps_nevpt2}

\subsubsection{Overview of the algorithm}
We now describe the implementation of t-NEVPT2 for the MPS reference wavefunctions, which we denote as \tmpsnevpt. 
To aid the discussion, we first rewrite the t-NEVPT2 energy in a compact form
\begin{align}
	\label{eq:e_general}
	E^{(2)} = E^{(2)}_{ext} + \sum_{[i]} E_{act}^{[i]} \ ,
\end{align}
where we combine terms involving the core and external orbitals
\begin{align}
	\label{eq:e_ext}
	E^{(2)}_{ext} 
	&= \frac{1}{4}\sum_{ijab} \frac{\vv{ij}{ab}\vv{ab}{ij}}{\e{i}+\e{j}-\e{a}-\e{b}}
	+\sum_{ia} \frac{\th{i}{a}\th{a}{i}}{\e{i}-\e{a}} \notag \\
	&+2\sum_{ixya} \frac{\th{a}{i}\vv{ix}{ay}\pdm{y}{x}}{\e{i}-\e{a}} \ 
\end{align}
and introduce a shorthand notation $E_{act}^{[i]}$ for the active-space components of the energy contributions $E^{[i]}$ in \cref{eq:e_contributions} ($i\in\{+1$, $-1$, $+2$, $-2$, $+1'$, $-1'$, $0'\}$). Each component $E_{act}^{[i]}$ can be expressed in the general form:
\begin{align}
	\label{eq:e_act}
	E_{act}^{[i]} = -\int_{0}^{\infty}  \boldsymbol{\epsilon^{[i]}}(\tau) \mathbf{G}\boldsymbol{^{[i]}}(\tau) \, \mathrm{d}\tau \ ,
\end{align}
where $\mathbf{G}\boldsymbol{^{[i]}}(\tau)$ is the Green's function tensor and $\boldsymbol{\epsilon^{[i]}}(\tau)$ is the integral prefactor tensor that can be computed using the one- and two-electron integrals and orbital energies ($\e{i}$ and $\e{a}$). Explicit equations for the elements of tensors $\boldsymbol{\epsilon^{[i]}}(\tau)$ and $\mathbf{G}\boldsymbol{^{[i]}}(\tau)$ are given in \cref{tab:e_gf_contr}.

\begin{table*}[t]
\begin{flushleft}
\captionsetup{justification=raggedright,singlelinecheck=false}
\caption{Equations for the elements of tensors $\boldsymbol{\epsilon^{[i]}}(\tau)$ and $\mathbf{G}\boldsymbol{^{[i]}}(\tau)$ used in the evaluation of the t-NEVPT2 active-space energy contributions $E_{act}^{[i]}$ in \cref{eq:e_act}.}
\label{tab:e_gf_contr}
{
\setstretch{1.5}
\begin{tabular}{C{1cm} C{7.5cm} C{7.5cm}}
\hline
\hline
$[i]$ & $\boldsymbol{\epsilon^{[i]}}(\tau)$ & $\mathbf{G}\boldsymbol{^{[i]}}(\tau)$  \\
\hline
$[+1]$		& $\epsilon_{x}^{y}(\tau) = \frac{1}{2} \sum_{ija} \vv{ij}{ay} \vv{ax}{ij} e^{(\e{i} + \e{j} - \e{a}) \tau}$		& $G_{x}^{y}(\tau) = \braket{\Psi_0| \a{x} (\tau) \c{y} |\Psi_0}$	 \\
$[-1]$		& $\epsilon_{y}^{x}(\tau) = \frac{1}{2} \sum_{iab} \vv{iy}{ab} \vv{ab}{ix} e^{(\e{i} - \e{a} - \e{b}) \tau} $ 	& $G_{y}^{x}(\tau) = \braket{\Psi_0| \c{x} (\tau) \a{y} |\Psi_0}$	 \\
$[+2]$		& $\epsilon_{xy}^{zw}(\tau) = \frac{1}{8} \sum_{ij} \vv{ij}{zw} \vv{xy}{ij} e^{(\e{i} + \e{j}) \tau} $			& $G_{xy}^{zw}(\tau) = \braket{\Psi_0| \a{x} (\tau) \a{y} (\tau) \c{w} \c{z} |\Psi_0}$	 \\
$[-2]$		& $\epsilon_{zw}^{xy}(\tau) = \frac{1}{8} \sum_{ab} \vv{zw}{ab} \vv{ab}{xy} e^{-(\e{a} + \e{b}) \tau}$	& $G_{zw}^{xy}(\tau) = \braket{\Psi_0| \c{x} (\tau) \c{y} (\tau) \a{w} \a{z} |\Psi_0}$	 \\
$[0']$		& $\epsilon_{yz}^{xw}(\tau) = \sum_{ia} \vv{iz}{aw} \vv{ay}{ix} e^{(\e{i} - \e{a})\tau} $				& $G_{yz}^{xw} (\tau) = \braket{\Psi_0| \c{x} (\tau) \a{y} (\tau) \c{w} \a{z} |\Psi_0} $	 \\
$[+1']$		& $\epsilon_{i}^{i} (\tau) = e^{\e{i} \tau}$												& $G_{i}^{i}(\tau) = \braket{\Psi_0| \hat{h}^{}_{i}  (\tau) \hat{h}^{\dag}_{i} |\Psi_0} $	 \\
$[-1']$		& $\epsilon_{a}^{a} (\tau) = e^{-\e{a} \tau} $											& $G_{a}^{a} (\tau) = \braket{\Psi_0| \hat{h}^{\dag}_{a}  (\tau) \hat{h}^{}_{a} |\Psi_0}$	 \\
\hline
\hline
\end{tabular}
}
\end{flushleft}
\end{table*}

The general t-NEVPT2 algorithm consists of three steps: (i) Green's function evaluation, (ii) computation of the energy as a function of imaginary time ($\tau$), and (iii) integration in imaginary time. In step (i), the elements of $\mathbf{G}\boldsymbol{^{[i]}}(\tau)$ are computed for a set of $\tau$ values (time steps, $\tau_k$). In step (ii), the prefactors $\boldsymbol{\epsilon^{[i]}}(\tau)$ are evaluated using the equations in \cref{tab:e_gf_contr}, and the energy contributions at each time step are computed as $E_{act}^{[i]}(\tau_k) = \boldsymbol{\epsilon^{[i]}}(\tau_k) \mathbf{G}\boldsymbol{^{[i]}}(\tau_k)$. For example, $E_{act}^{[-2]}(\tau_k)$ is computed as the dot product of two tensors: $\epsilon^{[-2]}(\tau_k)=\epsilon_{zw}^{xy}(\tau_k)=\frac{1}{8}\sum_{ab}\vv{zw}{ab}\vv{ab}{xy}e^{-(\e{a}+\e{b})\tau_k}$ and $G^{[-2]}(\tau_k)=G_{zw}^{xy}(\tau_k)=\braket{\Psi_0| \c{x} (\tau_k) \c{y} (\tau_k) \a{w} \a{z} |\Psi_0}$. Finally, in step (iii), each energy contribution $E_{act}^{[i]}$ is evaluated by fitting the computed values $E_{act}^{[i]}(\tau_k)$ to an exponential function $\sum_i\mathcal{E}(\tau)=a_ie^{-b_i\tau}$ and integrating the obtained result. We note that the above algorithm outline can be used to implement t-NEVPT2 for reference wavefunctions in any representation (e.g., determinant-based\cite{Sokolov:2016p064102} or MPS-based). Only the details of step (i) depend on the explicit form of $\ket{\Psi_0}$. In \cref{sec:t_mps_nevpt2_gf}, we will describe how imaginary-time Green's functions $\mathbf{G}\boldsymbol{^{[i]}}(\tau)$ are evaluated in the MPS representation.

\subsubsection{Green's function evaluation}
\label{sec:t_mps_nevpt2_gf}

To organize our MPS implementation of the imaginary-time Green's functions, we express the elements of $\mathbf{G}\boldsymbol{^{[i]}}(\tau)$ (\cref{tab:e_gf_contr}) in the general form $G_{\mu\nu}^{[i]}(\tau)=\braket{\Phi_{\mu}^{[i]}(\tau)|\Phi_{\nu}^{[i]}}$, where $\ket{\Phi_{\nu}^{[i]}}$ is obtained by applying all creation and annihilation operators at $\tau = 0$ on $\ket{\Psi_0}$ (e.g., $\a{x}\ket{\Psi_0}$, $\c{x}\a{y}\ket{\Psi_0}$), while the application of operators at $\tau=\tau'$ yields $\ket{\Phi_{\mu}^{[i]}(\tau')}$ (e.g., $\a{x}(\tau')\ket{\Psi_0}$, $\c{x}(\tau')\a{y}(\tau')\ket{\Psi_0}$). In our notation, indices $\mu$ and $\nu$ run over the total number of states $\ket{\Phi_{\mu}^{[i]}}$ (i.e., $N_{act}$ for $\a{x}\ket{\Psi_0}$ or $N^2_{act}$ for $\c{x}\a{y}\ket{\Psi_0}$). 

We begin the computation of $\mathbf{G}\boldsymbol{^{[i]}}(\tau)$ by optimizing the zeroth-order MPS wavefunction $\ket{\Psi_0}$ using the DMRG algorithm with  bond dimension $M_0$. To evaluate the elements of $\mathbf{G}\boldsymbol{^{[i]}}(\tau)$
we employ the algorithm below. 
\begin{enumerate}
\item {\it Compute initial wavefunctions} $\ket{\Phi_{\mu}^{[i]}}$ for every $\mu$. Each wavefunction $\ket{\Phi_{\mu}^{[i]}}$ is computed as an individual MPS and stored on disk. Evaluation of $\ket{\Phi_{\mu}^{[i]}}$ ($[i]$ $\in$ $\{[+1]$, $[-1]$, $[+2]$, $[-2]$, $[0']\}$) requires
  applying the operator $\c{x}$ or $\a{x}$ on $\ket{\Psi_0}$ one or more times. The resulting MPS is of exactly
  the same bond dimension as the original MPS. 
  For $[i]$ $\in$ $\{[+1']$, $[-1']\}$, the wavefunctions $\ket{\Phi_{\mu}^{[+1']}}$ and $\ket{\Phi_{\mu}^{[-1']}}$ are computed by compressing the MPS form of $\hat{h}_i^\dag\ket{\Psi_0}$ and $\hat{h}_a\ket{\Psi_0}$, where
$\hat{h}_i^\dag$ and $\hat{h}_a$ are defined as in \cref{eq:h_three_particle}.
Since applying $\hat{h}_i^\dag$ or $\hat{h}_a$ involves a summation over the active-space labels $x$, the resulting MPS is of larger bond dimension than $\ket{\Psi_0}$.
We use variational compression to obtain $\ket{\Phi_{\mu}^{[+1']}}$ and $\ket{\Phi_{\mu}^{[-1']}}$ MPS by 
minimizing  $|| \ket{\Phi_i^{[+1']}} - \hat{h}_i^\dag\ket{\Psi_0}||$ and $|| \ket{\Phi_a^{[-1']}} - \hat{h}_a\ket{\Psi_0}||$, where $\ket{\Phi_i^{[+1']}}$ and $\ket{\Phi_a^{[-1']}}$
have bond dimension $M_1 > M_0$. 
To  maximize the efficiency of computing the overlaps appearing in the compression (e.g.\@ $\bra{\Phi_i^{[+1']}}\hat{h}_i^\dag\ket{\Psi_0}$)
  we use the standard DMRG formalism of ``normal''  and ``complementary'' operators, which requires building $\mathcal{O}(k)$ such operators from the left and right blocks.
In practice, we find that $M_1\approx2M_0$ usually provides a sufficiently accurate compression. Overall, this step of the \tmpsnevpt algorithm has $\mathcal{O}(N_{act}^2M_0^3)+\mathcal{O}(N_{ext} N_{act}^2M_1^2M_0)$ scaling for computing all $\ket{\Phi_{\mu}^{[i]}}$ wavefunctions with $[i]$ $\in$ $\{[+2]$, $[-2]$, $[0']\}$ and $[i]$ $\in$ $\{[+1']$, $[-1']\}$, respectively ($N_{ext}$ is the number of external orbitals). 
\item {\it Propagate wavefunctions} $\ket{\Phi_{\mu}^{[i]}}$ in imaginary time $\tau$ according to the time-dependent Schr\"odinger equation $\ket{\Phi_{\mu}^{[i]}(\tau)} = e^{-(\hat{H}_D-E_D)\tau}\ket{\Phi_{\mu}^{[i]}}$.
In this step, the algorithm we use is based on the time-step targeting time-dependent DMRG algorithm of Feiguin and White described in Ref.~\citenum{Feiguin:2005p020404}, with some small
  modifications.
Specifically, we use the embedded Runge-Kutta (ERK) (4,5) time-step algorithm,\cite{Press:2007} which allows to automatically control the time step $\delta\tau$ based on the error estimate of the fifth-order approximation of the propagator $e^{-(\hat{H}_D-E_D)\tau}$. 
In the ERK (4,5) method, the wavefunction at time $\tau'=\tau+\delta\tau$ is approximated to $\mathcal{O}((\delta \tau)^5)$ as 
\begin{align}
	\label{eq:erk_wfn}
	\ket{\Phi_{\mu}^{[i]}(\tau')}\approx\ket{\Phi_{\mu}^{[i]}(\tau)}+\sum_{n=1}^6c_n\ket{k_n(\delta\tau)} \ ,
\end{align}
where states $\ket{k_n(\delta\tau)}$ are obtained by successively applying the zeroth-order Hamiltonian six times: 
\begin{widetext}
\begin{align}
	\label{eq:erk_k}
	\ket{k_n(\delta\tau)}=-\delta\tau(\hat{H}_D-E_D)\left[\ket{\Phi_{\mu}^{[i]}(\tau)}+\sum_{m=1}^{n-1}b_{nm}\ket{k_{m}(\delta\tau)}\right] \ .
\end{align}
\end{widetext}
The coefficients $b_{nm}$ and $c_n$ are given in Ref.\@ \citenum{Press:2007}. In addition to the fifth-order approximation, \cref{eq:erk_wfn} can be used to compute the fourth-order estimate of $\ket{\Phi_{\mu}^{[i]}(\tau')}$, which we denote as $\ket{\Phi_{\mu}^{[i]}(\tau')[4]}$. This can be done by setting $c_n=c_n[4]$, where the fourth-order coefficients $c_n[4]$ can be found in Ref.\@ \citenum{Press:2007}.
This gives an estimate of the time-step error, $\Delta = \sqrt{\left|\left|\ket{\Phi_{\mu}^{[i]}(\tau')} - \ket{\Phi_{\mu}^{[i]}(\tau')[4]}\right|\right|}$.

As mentioned in \cref{sec:dmrg}, we can adapt the above general wavefunction propagation to the propagation of a MPS
within a sweep algorithm. This is the idea behind the time-step targeting \tddmrg.
The quantities in the above equations are then to be interpreted as applying to each site, i.e.\@
the wavefunction $\ket{\Phi_{\mu}^{[i]}}$ is now the site wavefunction, the states $\ket{k_n}$
are vectors in the renormalized site basis, and the Hamiltonian $\hat{H}_D$ is projected into the renormalized site basis.
The site error in the
propagator approximation is estimated from the site wavefunction as $\Delta_p$. 
As in Ref.~\citenum{Feiguin:2005p020404}, the site wavefunctions $\ket{\Phi_{\mu}^{[i]}(\tau)}$, $\ket{\Phi_{\mu}^{[i]}(\tau+\delta\tau/3)}$,
$\ket{\Phi_{\mu}^{[i]}(\tau+2 \delta\tau/3)}$, $\ket{\Phi_{\mu}^{[i]}(\tau+\delta \tau)}$ are averaged in the density matrix to construct
density matrix eigenvectors to update the site wavefunction, before proceeding to the next step in the sweep.

 We begin the time propagation with a small value of $\delta\tau\sim10^{-3}$ and determine the new time-step after
each propagation as $\delta\tau' = \min(2\times\delta\tau,\delta\tau_{emb})$, where $\delta\tau_{emb}=\delta\tau\left|\frac{\Delta E_{conv}}{\max_p(\{\Delta_p\})}\right|^{1/5}$ and $\Delta E_{conv}$ is the specified accuracy threshold. Note that if $\max_p(\{\Delta_p\})>\Delta E_{conv}$, the time step is decreased. In this case, we repeat the time propagation using the smaller time step. The computational cost of a single time step for all states $\ket{\Phi_{\mu}^{[i]}}$ has $\mathcal{O}(N_{act}^5M_0^3)+\mathcal{O}(N_{act}^6M_0^2)+\mathcal{O}(N_{ext} N_{act}^3M_1^3)+\mathcal{O}(N_{ext} N_{act}^4M_1^2)$ scaling. 
\item {\it Compute Green's function elements} at every time step $\tau_k$ as the overlap between two MPS $G_{\mu\nu}^{[i]}(\tau_k)=\braket{\Phi_{\mu}^{[i]}(\tau_k)|\Phi_{\nu}^{[i]}}$.
  The computational cost of computing all $G_{\mu\nu}^{[i]}(\tau_k)$ elements has $\mathcal{O}(N_{act}^5M_0^3)$ scaling.
\end{enumerate}
The cost of the \tmpsnevpt algorithm outlined above is dominated by the computation of the 2-GF, which requires time-propagation of $\mathcal{O}(N_{act}^2)$ MPS wavefunctions $\ket{\Phi_{\mu}^{[i]}}$ and evaluation of $\mathcal{O}(N_{act}^4)$ matrix elements $G_{\mu\nu}^{[i]}(\tau_k)$, leading to a total $\mathcal{O}(N_\tau N_{act}^5M_0^3)+\mathcal{O}(N_\tau N_{act}^6M_0^2)$ computational scaling, where $N_\tau$ $\sim$ 10-20 is the number of time steps. While this is less than the cost of computing the 4-RDM in DMRG, computing the 3-RDM, which has $\mathcal{O}(N_{act}^4M_0^3)+\mathcal{O}(N_{act}^6M_0^2)$ scaling, is of lower cost.\cite{Guo:2016p1583} 
The higher scaling of our 2-GF implementation is because each MPS $\ket{\Phi_{\mu}^{[i]}}$ is propagated in imaginary time independently,
while in the efficient 3-RDM implementation a common set of renormalized operators is reused to compute all elements of the density matrix together. However, the higher scaling of this particular implementation is not an intrinsic property of the time-dependent theory. In the determinant basis, the computation of the 2-GF and 3-RDM have the same computational scaling.
For MPS, it is possible to similarly devise an algorithm to compute the 2-GF with the same cost as the 3-RDM
by propagating multiple $\ket{\Phi_{\mu}^{[i]}}$ wavefunctions using the same renormalized basis (similar to
performing a state-averaged DMRG optimization, and related to the algorithm used in Ref.~\citenum{Roemelt:2016p204113}). 
An advantage of our current \tmpsnevpt implementation, however, is that it is easily parallelized by separating the correlation energy contributions into $\mathcal{O}(N_{act}^2)$ components:
\begin{align}
	\label{eq:e_mu_components}
	E_{act}^{[i]} = -\sum_\mu\int_{0}^{\infty} \sum_\nu\epsilon_{\mu\nu}^{[i]}(\tau) G_{\mu\nu}^{[i]}(\tau)  \, \mathrm{d}\tau = \sum_\mu E_\mu^{[i]} \ ,
\end{align}
where each component $E_\mu^{[i]}$ can be evaluated independently with $\mathcal{O}(N_\tau N_{act}^3M_0^3)+\mathcal{O}(N_\tau N_{act}^4M_0^2)$ cost. In addition, as we will demonstrate in \cref{sec:results}, the energy terms that depend on the 2-GF converge very quickly with increasing $M_0$ and do not require a large bond dimension.

\subsubsection{Spin-adaptation}
\label{sec:t_mps_nevpt2_spin_adaptation}
In \cref{sec:t_mps_nevpt2_gf} we discussed the evaluation of the $G_{\mu\nu}^{[i]}(\tau)$ matrix elements in terms of creation and annihilation operators $\c{p}(\tau)$ and $\a{p}(\tau)$. In a non-spin-adapted DMRG algorithm, $a^\dag_{}$ and $a$ are spin-orbital operators and the index $p$ corresponds to a spatial orbital with a particular spin. For example, for  $[i]$ $=$ $[-2]$, the spin-orbital 2-GF has the following form: $G_{z_\lambda w_\kappa}^{x_\rho y_\sigma}(\tau)=\braket{\Psi_0| \c{x_\rho} (\tau) \c{y_\sigma} (\tau) \a{w_\kappa} \a{z_\lambda} |\Psi_0}$, where we use the Roman characters $x,y,w,z$ to denote spatial orbitals and Greek characters $\rho,\sigma,\kappa,\lambda$ to denote the spin of these orbitals. Here, $\a{w_\kappa}$ and $\a{z_\lambda}$ are simple operators, and application of a pair of operators $\a{w_\kappa} \a{z_\lambda} \ket{\Psi_0}$ results in a single MPS wavefunction $\ket{\Phi_{w_\kappa z_\lambda}^{[-2]}}$. 

In a spin-adapted DMRG implementation,\cite{Sharma:2012p124121} $a^\dag_{}$ and $a$ refer to spin tensor creation and annihilation operators, and strings
of these operators generate multiple eigenstates of the $\hat{S}^2$ operator when acting on a reference state. As a result, expectation values of spin tensor operators depend on the spin quantum numbers, e.g.\@ $\braket{\Psi_0| \{[\c{x} (\tau) \c{y} (\tau)]^{S_1}_{M_1} [\a{w} \a{z}]^{S_2}_{M_2}\}^{S_3}_{M_3} |\Psi_0}$, where we use brackets to denote the coupling of spins $S_1$, $S_2$, and $S_3$ for different tensor products. As an example, we consider the evaluation of $G_{\mu\nu}^{[-2]}(\tau)$ for a reference state $\ket{\Psi_0}$ with $S = 0$. (Generalization to reference states with $S \ne 0$ is straightforward and is simplified using singlet embedding).\cite{Tatsuaki:2000p3199,Sharma:2012p124121} Applying a pair of tensor operators on the reference state $[\a{w} \a{z}]^{S_1}_{M_1}\ket{\Psi_0}$ results in two sets of eigenstates $\ket{{\Phi_{wz}^{[-2]}}_{M_1}^{S_1}}$ with $S_1$ $=$ $0,1$. The wavefunctions $\ket{{\Phi_{wz}^{[-2]}}_{M_1}^{S_1}}$ are propagated in imaginary time for each value of $S_1$, and the matrix elements are computed using Clebsch-Gordan coefficients $c_{M_1,M_2,M_3}^{S_1,S_2,S_3}$ as $(G_{zw}^{xy})^{S_1}(\tau)=\sum_{M_1}(-1)^{S_1+M_1}c_{M_1,-M_1,0}^{S_1,S_1,0}\braket{\Psi_0| [\c{x} (\tau) \c{y} (\tau)]^{S_1}_{M_1} [\a{w} \a{z}]^{S_1}_{M_1}|\Psi_0}$, where we used the fact that $S_1=S_2$ and $S_3=0$ for a closed-shell $\ket{\Psi_0}$. Expanding the spin tensor operators $[\a{w} \a{z}]^{S_1}$ as a combination of spin-orbital operators $\a{w_\rho}$ and $\a{z_\sigma}$ leads to a system of linear equations that can be solved to obtain the spin-orbital 2-GF elements $G_{z_\lambda w_\kappa}^{x_\rho y_\sigma}(\tau)$ from the matrix elements $(G_{zw}^{xy})^{S_1}(\tau)$ (see the Appendix A for explicit equations). 

\section{Computational Details}
\label{sec:comp_details}

All DMRG computations for the reference wavefunctions, reduced density matrices, and Green's functions were performed using the \textsc{Block} program.\cite{Sharma:2012p124121} The \tmpsnevpt and \scmpsnevpt correlation energies were computed using \textsc{Pyscf},\cite{Sun:ArXiv08223} through its interface with \textsc{Block}. For all NEVPT2 computations, the DMRG reference wavefunctions were fully optimized with respect to active-external orbital rotations using the \dmrgscf algorithm implemented in \textsc{Pyscf}. We denote active spaces used in \dmrgscf as ($n$e, $m$o), where $n$ is the number of active electrons and $m$ is the number of orbitals. We used tight convergence parameters for the energy in the DMRG sweeps (10$^{-9}$ \eh) and orbital optimization iterations (10$^{-6}$ \eh). In \tmpsnevpt, the accuracy of time propagation was controlled by setting $\Delta E_{conv}=10^{-4}$ \eh for all systems but the N$_2$ molecule, where the tighter threshold $\Delta E_{conv}=10^{-6}$ \eh was used (see \cref{sec:t_mps_nevpt2_gf} for details). Specific details pertaining to our computations of the {\it all-trans} polyenes are described in \cref{sec:polyenes}.

As described in \cref{sec:t_mps_nevpt2}, the bond dimensions necessary to represent the reference MPS ($M_0$) and the compressed MPS ($M_1$) are usually different ($M_0<M_1$). However, to simplify the discussion, in our computations we set $M_0$ and $M_1$ to the same value, and whenever we refer to bond dimension $M$ we imply that $M_0=M_1=M$.

\section{Results}
\label{sec:results}

\subsection{Analysis of numerical accuracy: N$_2$ dissociation}
\label{sec:numerical_accuracy}

We first investigate the numerical accuracy of our algorithm by computing the errors of \tmpsnevpt as a function of the bond dimension relative to the uncontracted NEVPT2 energies computed using our t-NEVPT2/CASSCF implementation.\cite{Sokolov:2016p064102} For this purpose, we evaluate the \tmpsnevpt energies for a range of geometries along the ground-state potential energy curve (PEC) of the \ce{N2} molecule. At each geometry, we optimize the molecular orbitals using the CASSCF method in the (10e,10o) active space and perform a single DMRG computation using these orbitals to obtain a MPS wavefunction with 
a large bond dimension $M'$ = 1000. We subsequently compress this reference wavefunction down to an MPS with a smaller bond dimension ($M$ = 50, 100, 200) to be used in the NEVPT2 computation. 
In this section, we use Dunning's cc-pVQZ basis.\cite{Dunning:1989p1007}

\begin{figure}[t!]
   \includegraphics[width=0.45\textwidth]{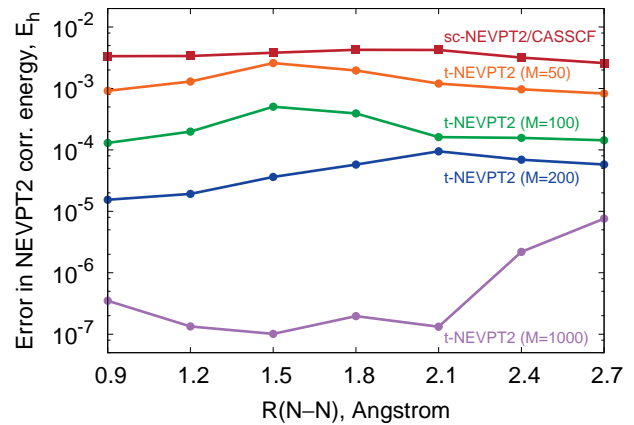}
   \captionsetup{justification=raggedright,singlelinecheck=false}
   \caption{Error in the \tmpsnevpt correlation energy relative to that of the uncontracted NEVPT2 theory for \ce{N2} as a function of the N--N bond distance and the DMRG bond dimension $M$. Computations employed the (10e, 10o) active space and the cc-pVQZ basis set. Also shown are the strong contraction errors computed as the difference between sc-NEVPT2 and t-NEVPT2 based on the CASSCF reference.}
   \label{fig:N2_diff}
\end{figure}

\cref{fig:N2_diff} plots the \tmpsnevpt correlation energy errors for different values of the bond dimension $M$, as well as the error in sc-NEVPT2/CASSCF due to the strong contraction approximation. Notably, even with a small $M$ = 50, the error from finite bond dimension in \tmpsnevpt is significantly smaller than the contraction error of sc-NEVPT2 for all geometries along the \ce{N2} potential energy curve (PEC). As the bond dimension $M$ increases, the \tmpsnevpt errors decrease, becoming smaller than 0.1 \meh at $M$ = 200 for all energy points. As can be seen from \cref{fig:N2_diff}, the convergence of \tmpsnevpt energies is not monotonic along the dissociation curve. In particular, in the range of short bond distances ($r$ $<$ 2.1 \AA) the correlation energies converge significantly faster to the exact (uncontracted) NEVPT2 energies than the energies at the dissociation limit ($r$ $\ge$ 2.1 \AA). This result is consistent with the fact that
both the zeroth-order and first-order wavefunctions  at longer bond distances contain contributions from a larger number of determinants than at the shorter bond distances. To illustrate this, we computed the \tmpsnevpt energies using a large bond dimension $M$ = 1000 (\cref{fig:N2_diff}). For all geometries with $r$ $\le$ 2.1 \AA, our \tmpsnevpt algorithm achieves an accuracy of better than $10^{-6}$ \eh, while errors in the range of $\sim$ $10^{-5}$ \eh still remain in the dissociation limit. The slow convergence of correlation energies with bond dimension near the dissociation limit has also been observed in Sharma and Chan's MPS-PT2 study of  \ce{N2} dissociation using the cc-pVDZ basis set and the (10e,8o) active space.\cite{Sharma:2014p111101}

\begin{figure}[t!]
   \subfloat[]{\label{fig:N2_diff_2e}\includegraphics[width=0.45\textwidth]{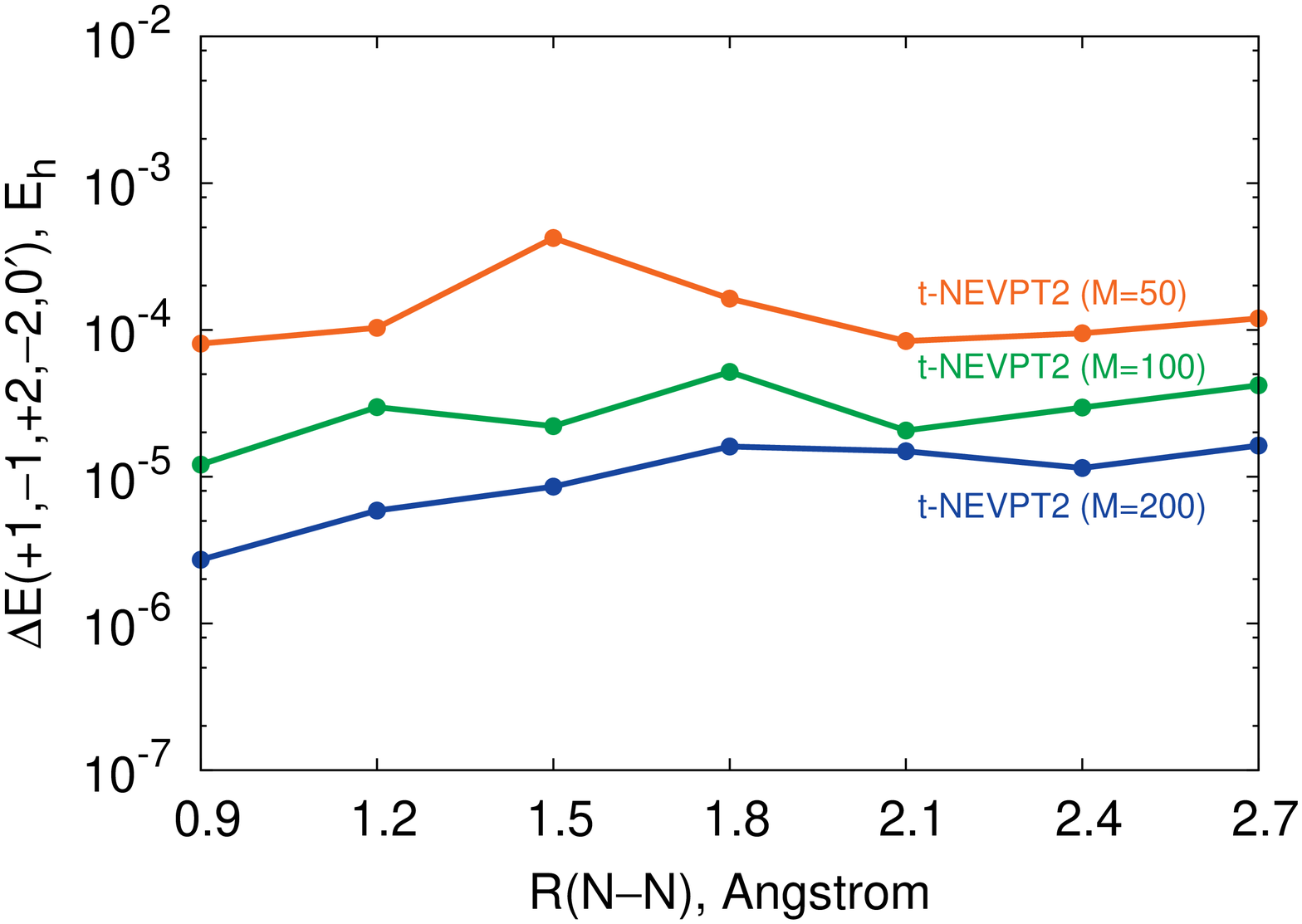}} \quad
   \subfloat[]{\label{fig:N2_diff_3e}\includegraphics[width=0.45\textwidth]{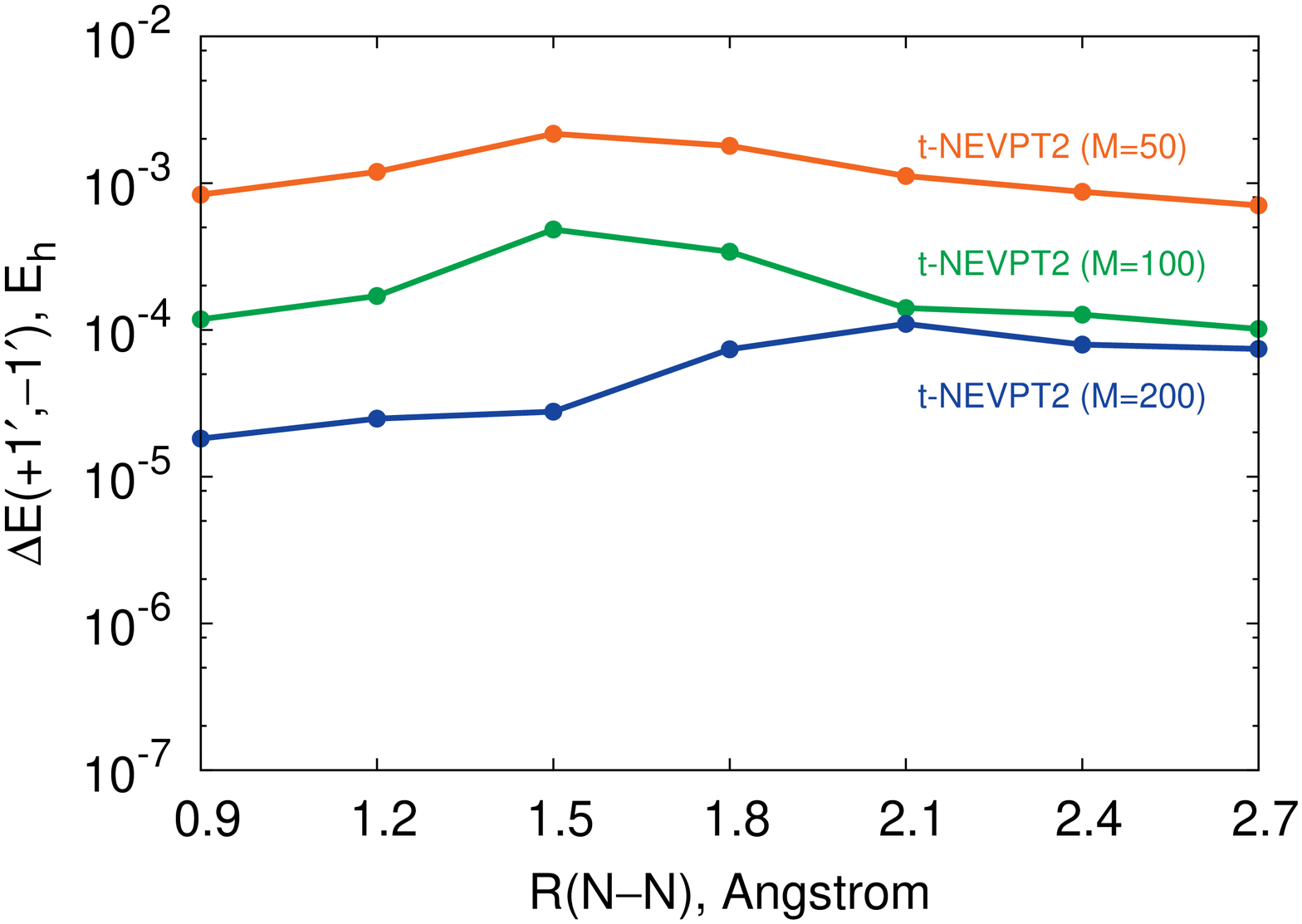}} 
   \captionsetup{justification=raggedright,singlelinecheck=false}
   \caption{Error in the \tmpsnevpt correlation energy contributions for \ce{N2} as a function of the N--N bond distance and the DMRG bond dimension $M$. Plot (a) shows the errors for $E_{act}^{[i]}$ ($[i]$ $\in$ $\{[+1]$, $[-1]$, $[+2]$, $[-2]$, $[0']\}$) components that depend on the 1- and 2-GF, while plot (b) presents the errors for $E_{act}^{[i]}$ ($[i]$ $\in$ $\{[+1']$, $[-1']\}$) that depend on the contracted 3-GF. Computations employed the (10e, 10o) active space and the cc-pVQZ basis set.}
   \label{fig:N2_diff_components}
\end{figure}

We now analyze the errors in the \tmpsnevpt correlation energy from two different contributions: (i) the sum of active-space energy components that depend on the 1- and 2-GF $\sum_{[i]}E_{act}^{[i]}$ ($[i]$ $\in$ $\{[+1]$, $[-1]$, $[+2]$, $[-2]$, $[0']\}$), and (ii) the sum of energy terms that depend on the contracted 3-GF ($E_{act}^{[+1']}$ + $E_{act}^{[-1']}$). These errors are plotted in \cref{fig:N2_diff_2e,fig:N2_diff_3e} for different values of $r$ and $M$. Between the two energy contributions, the contribution (i) from the 1- and 2-GF exhibits significantly faster convergence with increasing $M$ than  contribution (ii) from the contracted 3-GF. At $M$ = 50, contribution (i) shows errors of $\sim$ $10^{-4}$ \eh for a large range of geometries, while the errors in contribution (ii) are an order of magnitude larger. Comparing \cref{fig:N2_diff_2e,fig:N2_diff_3e}, we observe that the energy components (i) and (ii) achieve similar accuracy when computed with bond dimensions of  $M$ and $2M$, respectively. In fact, for a specified $M$, the errors in the total \tmpsnevpt correlation energy are dominated by the errors in contribution (ii) ({\it cf.}\@ \cref{fig:N2_diff} and \cref{fig:N2_diff_3e}). As discussed in \cref{sec:t_mps_nevpt2_gf}, the observed slower convergence of contribution (ii) is the result of the MPS compression used to evaluate $E_{act}^{[+1']}$ and $E_{act}^{[-1']}$, which requires a larger bond dimension than the one used in the optimization of the reference MPS for an accurate representation. Since the two contributions can be evaluated separately, we recommend to compute contribution (i) using a small bond dimension $M$ and to use a larger bond dimension $\sim$ $2M$ to obtain contribution (ii).

\subsection{Chromium dimer}
\label{sec:cr2}
The chromium dimer (\ce{Cr2}) is a notoriously challenging system for multi-reference methods. Computational studies\cite{Andersson:1994p391,Roos:1995p215,Andersson:1995p212,Stoll:1996p793,Dachsel:1999p152,Roos:2003p265,Celani:2004p2369,Angeli:2006p054108,Ruiperez:2011p1640,Muller:2009p12729,Kurashige:2011p094104,Purwanto:2015p064302,Sokolov:2016p064102,Guo:2016p1583} have demonstrated that including the dynamic correlation in combination with the multi-configurational description of static correlation and large basis sets is crucial to properly describe the ground-state PEC of this molecule. In 2011, Kurashige and Yanai showed that the \ce{Cr2} dissociation curve can be accurately described by combining the CASPT2 variant of internally-contracted perturbation theory with DMRG in a large (12e,28o) active space,\cite{Kurashige:2011p094104} which included the $3d$, $4d$, $4s$, and $4p$ orbitals of the Cr atoms. Very recently (2016), Guo {\it et al.}\@ computed \ce{Cr2} PEC using a combination of sc-NEVPT2 and DMRG (\scnevpt) with the (12e,22o) active space ($3d$, $4d$, and $4s$ orbitals of Cr).\cite{Guo:2016p1583} At the complete basis set limit (CBS), the \ce{Cr2} PEC from \scnevpt agrees well with the dissociation curve from the experiment, yielding spectroscopic parameters for the equilibrium bond length $r_e$ = 1.656 \AA and a binding energy $D_e$ = 1.432 eV that are close to the experimental $r_e^{exp}$ = 1.679 \AA and $D_e^{exp}$ = 1.47 $\pm$ 0.1 eV.\cite{Casey:1993p816} However, the \ce{Cr2} binding energy in sc-NEVPT2 can be significantly affected by the strong contraction approximation, as we demonstrated in our t-NEVPT2 study of the \ce{Cr2} PEC using the (12e,12o) active space.\cite{Sokolov:2016p064102} In this work, we recompute the \ce{Cr2} equilibrium binding energy in the large (12e,22o) active space using our \tmpsnevpt algorithm.

We obtained the MPS reference wavefunction as described in the \scnevpt study by Guo {\it et al.}\cite{Guo:2016p1583} In short, we used the (12e,22o) active space and optimized the molecular orbitals using the \dmrgscf algorithm with $M'$ = 1000 (no frozen core). We then performed a DMRG computation with $M'$ = 6000 using the optimized orbitals to obtain an accurate reference wavefunction. To compute imaginary-time Green's functions, the reference MPS with the large bond dimension $M'$ was compressed down to an MPS with a smaller bond dimension $M$. We computed each \tmpsnevpt energy contribution $E_{act}^{[i]}$ by starting with $M=500$ and increasing $M$ until convergence. To account for the scalar relativistic effects, we used the cc-pwCV5Z-DK basis set\cite{Balabanov:2005p064107} combined with the spin-free one-electron variant of the X2C Hamiltonian.\cite{Liu:2010p1679,Saue:2011p3077,Peng:2012p1081} Since  strong contraction reduces the binding energy\cite{Sokolov:2016p064102} in \ce{Cr2}, the CBS-extrapolated equilibrium bond length $r_e$ computed using the uncontracted \tmpsnevpt method is expected to be somewhat shorter than $r_e$ = 1.656 \AA obtained from \scnevpt. Based on the results of our preliminary study,\cite{Sokolov:2016p064102} we estimate $r_e$ obtained from the uncontracted theory to be $\sim$ 0.005 \AA shorter, and use the bond distance $r$(Cr--Cr) = 1.650 \AA in our computations.

\begin{table*}[t]
\begin{flushleft}
\captionsetup{justification=raggedright,singlelinecheck=false}
\caption{\tmpsnevpt correlation energy contributions $E_{act}^{[i]}$ for Cr$_2$ at 1.650 \AA computed for different values of the bond dimension $M$. Computations employed the (12e,22o) active space and the cc-pwCV5Z-DK basis set. Also shown are strong contraction errors $\Delta E_{contr}$ in the correlation energy for Cr$_2$, the Cr atom, and the Cr$_2$ binding energy ($D_e$), evaluated as the difference between the \scnevpt and \tmpsnevpt energies.}
\label{tab:Cr2}
{
\setstretch{1.5}
\begin{tabular}{C{1.25cm} C{2cm} C{2cm} C{2cm} C{2cm} C{2cm} C{2cm} C{2cm}}
\hline
\hline
$M$ & $E_{act}^{[+1]}$ &  $E_{act}^{[-1]}$  &  $E_{act}^{[+2]}$  & $E_{act}^{[-2]}$  &  $E_{act}^{[0']}$  &  $E_{act}^{[+1']}$  &  $E_{act}^{[-1']}$ \\
\hline
500	& $-$0.1632	& $-$0.2357	& $-$0.0939	& $-$0.1140	& $-$1.6425	& $-$0.0282	& $-$0.0416	\\
800	& $-$0.1633	& $-$0.2357	& $-$0.0940	& $-$0.1140	& $-$1.6422	& $-$0.0286	& $-$0.0431	\\
1200	& 	& 	& 	& 	& 	& $-$0.0288	& $-$0.0441	\\
1600	& 	& 	& 	& 	& 	& $-$0.0290	& $-$0.0446	\\
2000	& 	& 	& 	& 	& 	& $-$0.0291	& $-$0.0449	\\
\multicolumn{8}{c}{
$\Delta E_{contr} (\mathrm{Cr_2}) = E^{(2)}_{\mathrm{sc-NEVPT2}} - E^{(2)}_{ext} - \sum_{[i]} E_{act}^{[i]}$
= $-$1.6351 $-$ 0.6801 + 2.3232 = 0.0080 \eh
} \\
\multicolumn{8}{c}{
$\Delta E_{contr} (\mathrm{Cr}) = E^{(2)}_{\mathrm{sc-NEVPT2}}(\mathrm{Cr}) - E^{(2)}_{\mathrm{t-NEVPT2}}(\mathrm{Cr})$
= $-$0.7758 $+$ 0.7781 = 0.0023 \eh
} \\
\multicolumn{8}{c}{
$\Delta E_{contr} (D_e)$ = $\Delta E_{contr} (\mathrm{Cr_2})$ $-$ $2\times\Delta E_{contr} (\mathrm{Cr})$ = 0.0034 \eh = {\bf 0.09 eV}
} \\
\hline
\hline
\end{tabular}
}
\end{flushleft}
\end{table*}

\cref{tab:Cr2} presents \tmpsnevpt correlation energy contributions computed for different values of the bond dimension $M$. Increasing $M$ from 500 to 800 does not significantly affect the energy components $E_{act}^{[i]}$ that depend on the 1-GF ($[i]$ $\in$ $\{[+1]$, $[-1]\}$) and 2-GF ($[i]$ $\in$ $\{[+2]$, $[-2]$, $[0']\}$). Out of five of these contributions, the largest change of +0.3 \meh is observed for $E_{act}^{[0']}$, while each of the other four energy terms changes by less than $-$0.1 \meh. These changes largely cancel each other, leading to a total energy difference of $\sim$ 0.1 \meh. Thus, we estimate the total error in the sum of the 1- and 2-GF energy contributions to be less than 0.1 \meh. However, the remaining energy components $E_{act}^{[+1']}$ and $E_{act}^{[-1']}$ that depend on the contracted 3-GF exhibit slower convergence with $M$, changing by $-$1.9 \meh for $M$ = 500 $\rightarrow$ 800 and $-$1.2 \meh for $M$ = 800 $\rightarrow$ 1200 (\cref{tab:Cr2}). We increased the bond dimension up to $M$ = 2000, where the remaining error in $E_{act}^{[+1']}$ and $E_{act}^{[-1']}$ is estimated to be less than 0.4 \meh ($<$ 0.01 eV). The resulting active-space energy contributions $E_{act}^{[i]}$ can be used to compute the \ce{Cr2} binding energy for the incomplete cc-pwCV5Z-DK basis set as $D_e$ =  $E^{(2)}_{ext}$ $+$ $\sum_{[i]} E_{act}^{[i]}$ $-$ $E^{(2)}_{\mathrm{t-NEVPT2}}(\mathrm{Cr})$, where $E^{(2)}_{ext}$ is the \ce{Cr2} correlation energy from the external space (\cref{eq:e_ext}) and $E^{(2)}_{\mathrm{t-NEVPT2}}(\mathrm{Cr})$ is the correlation energy for an isolated Cr atom. To estimate $D_e$ at the CBS limit, we first compute the error of strong contraction in the binding energy $\Delta E_{contr} (D_e)$ as  shown in \cref{tab:Cr2}, where $E^{(2)}_{\mathrm{sc-NEVPT2}}$ is the \scnevpt correlation energy for \ce{Cr2} reported by Guo {\it et al.}\@ ($M$ = 1200),\cite{Guo:2016p1583} while $E^{(2)}_{\mathrm{sc-NEVPT2}}(\mathrm{Cr})$ and $E^{(2)}_{\mathrm{t-NEVPT2}}(\mathrm{Cr})$ are the correlation energies for a Cr atom computed using the (6e,11o) CASSCF reference. Assuming that the contraction error does not change significantly from the incomplete (cc-pwCV5Z-DK) to the complete basis set, the CBS-extrapolated \ce{Cr2} binding energy at the \tmpsnevpt level of theory can be estimated as $(D_e)_{\mathrm{t-NEVPT2}}^{\mathrm{CBS}}$ $\approx$ $(D_e)_{\mathrm{sc-NEVPT2}}^{\mathrm{CBS}}$ + $\Delta E_{contr} (D_e)$, where we use $(D_e)_{\mathrm{sc-NEVPT2}}^{\mathrm{CBS}}$ = 1.43 eV from Ref.\@ \citenum{Guo:2016p1583}. Although the resulting binding energy $(D_e)_{\mathrm{t-NEVPT2}}^{\mathrm{CBS}}$ = 1.52 eV is 0.09 eV larger than $(D_e)_{\mathrm{sc-NEVPT2}}^{\mathrm{CBS}}$, it is still in a good agreement with the experimental binding energy of 1.47 $\pm$ 0.1 eV reported by Casey and Leopold.\cite{Casey:1993p816}

\subsection{Singlet excited states in {\it all-trans} polyenes}
\label{sec:polyenes}

\subsubsection{Background}
Conjugated polyenes are important model systems for understanding properties of organic materials (such as polyacetylene), as well as pigments involved in photosynthesis and vision (e.g., carotenoids or retinals). The key to the functionality of these molecules is the ability to absorb and efficiently transfer energy via the low-lying $\pi$-$\pi^*$ excited states of the conjugated backbone. One of the simplest examples of polyene systems are the unsubstituted {\it all-trans} polyenes that consist of alternating single and double carbon bonds arranged in a linear chain (C$_n$H$_{n+2}$). The excited states of these molecules are labeled by the symmetry of the $C_{2h}$ point group ($A_g$, $A_u$, $B_g$, and $B_u$), as well as an additional $+/-$ label due to their approximate particle-hole symmetry, which is typically used to distinguish excited states with mainly ionic ($+$ states) or mainly covalent ($-$ states) excitation character.\cite{Pariser:1956p250,Tavan:1986p6602}

The electronic spectra of {\it all-trans} polyenes have been the subject of many experimental and computational studies.\cite{Tavan:1986p6602,Tavan:1987p4337,Nakayama:1998p157,Kurashige:2004p425,Ghosh:2008p144117,Zgid:2009p194107,Angeli:2011p184302,Watson:2012p4013,Ronca:2014p4014,Hudson:1973p4984,Sashima:1999p187,Cerullo:2002p2395,Furuichi:2002p547,Doering:1981p2477,Doering:1980p3617,Heimbrook:1981p4338,Heimbrook:1984p1592,Leopold:1984p4210,Petek:1993p3777,Kohler:1988p5422} Of particular interest are the excitation energies and ordering of the low-lying singlet electronic states. While the symmetry of the singlet ground state is known to always be $A_g^-$, the experimental and theoretical characterization of the low-lying singlet excited states has been very challenging. In shorter polyenes (C$_n$H$_{n+2}$, $n$ $\le$ 8), it has been established that the two lowest-energy singlet transitions are the dipole-allowed excitation to the ionic \bup state and the dipole-forbidden excitation to the \ag{2} state. Despite the fact that the allowed transition \ag{1} $\rightarrow$ \bup has primarily HOMO $\rightarrow$ LUMO excitation character, fluorescence studies suggest that,
starting from octatetraene (\ce{C8H10}),\cite{Hudson:1973p4984} the lowest-energy singlet excitation corresponds to a forbidden transition, such as the \ag{1} $\rightarrow$ \ag{2} transition. Recent high-level theoretical studies of butadiene (\ce{C4H6}) predict the bright \bup state vertical excitation to be $\sim$ 0.2 eV lower in energy than the excitation to the dark \ag{2} state,\cite{Watson:2012p4013} whereas the two vertical transitions are predicted to be almost degenerate in \ce{C8H10}.\cite{Angeli:2011p184302} In longer polyenes, the low-energy electronic spectrum is more complicated, involving additional dark states \bum and \ag{3} with energies close to \bup and \ag{2}. These dark states have been observed experimentally in {\it all-trans-}carotenoids with $n$ = 18--26 $\pi$-electrons.\cite{Sashima:1999p187,Cerullo:2002p2395,Furuichi:2002p547} 

Early computational studies of long polyenes have been primarily based on model Hamiltonians, such as the Pariser-Parr-Pople (PPP) and Hubbard models.\cite{Tavan:1986p6602,Tavan:1987p4337} In 1986, Tavan and Schulten\cite{Tavan:1986p6602} performed multi-reference configuration interaction (MRCI) computations using the PPP model, which suggested that at $n$ $=$ 10 the \bum state becomes the second excited state, leading to the \ag{1} $<$ \ag{2} $<$ \bum $<$ \bup $<$ \ag{3} ordering of states for longer polyenes. These computational results were recently reassessed using a combination of MRCI and the semi-empirical OM2 method,\cite{Schmidt:2012p124309} which predicted the inversion of the \bum and \bup transitions at  $n$ $=$ 14. 

In 1998, Hirao and co-workers performed the first {\it ab initio} computations of the vertical excitation energies in small and medium-size polyenes ($n$ $\le$ 10) using multi-reference perturbation theory (MRMP) based on the CASSCF reference wavefunctions.\cite{Nakayama:1998p157} In a later study, Kurashige {\it et al.}\@ used a combination of MRMP and CASCI with the (10e,10o) active space ({\it incomplete} $\pi$-valence space) for polyene series up to $n$ = 28.\cite{Kurashige:2004p425} Their computations also suggested that the \bum and \bup vertical transitions cross at $n$ = 14 and predicted a crossing of \ag{3} and \bup vertical excitations at $n$ = 22. In 2008, Ghosh {\it et al.} performed a \dmrgscf study of polyenes with 8 $\le$ $n$ $\le$ 24 using the {\it complete} $\pi$-valence active space ($n$e,$n$o).\cite{Ghosh:2008p144117} Although in this study the ionic \bup state was not investigated, the authors demonstrated that the self-consistent optimization of orbitals in the DMRG computations of these systems is very important in achieving reasonable agreement with the experiment for the covalent \ag{2}, \bum, and \ag{3} states. The work by Ghosh {\it et al.}, however, did not incorporate dynamic correlation. Although it is generally believed that the dynamic correlation mainly affects the ionic \bup state, excitation energies of the covalent states computed using \dmrgscf overestimate those obtained from the experiment.\cite{Ghosh:2008p144117} Using our \scmpsnevpt and \tmpsnevpt methods, we are now in the position to investigate the importance of dynamic correlation effects on the electronic excitations in {\it all-trans} polyenes in combination with the {\it complete} $\pi$-valence space description of static correlation.

\subsubsection{Details of computations}
The ground-state molecular geometries of the {\it all-trans} polyenes C$_n$H$_{n+2}$ ($n$ = 4, 8, 16, and 24) were optimized using the B3LYP method\cite{Becke:1993p5648,Lee:1988p785} implemented in the \textsc{Psi4} program.\cite{Turney:2012p556} Geometry optimizations were constrained to have $C_{2h}$ symmetry. For all polyenes, we used the cc-pVDZ basis set.\cite{Dunning:1989p1007} In addition, to study the effect of the basis set on the excitation energies, we also employed the larger aug-cc-pVTZ basis for the smaller polyenes ($n$ = 4 and 8).

At each optimized geometry, the reference MPS wavefunctions were computed using the \dmrgscf algorithm for the complete $\pi$-valence active space ($n$e,$n$o). To construct this active space, we first performed a self-consistent field (SCF) computation to obtain a set of canonical Hartree-Fock orbitals. The occupied subset of the SCF orbitals was used to generate an orthonormal set of intrinsic atomic orbitals (IAO),\cite{Knizia:2013p4834} from which the $2p_{z}$ orbitals of carbon atoms ($z$ being the $C_2$ axis) were easily identified and used as the initial active space orbitals. The initial core and external orbitals were obtained by projecting the active-space orbitals out of the occupied and virtual subspaces of the SCF orbitals, respectively, followed by their symmetric orthogonalization. These initial orbitals were subsequently optimized using the state-averaged \dmrgscf algorithm with $M'$ = 250, averaging over the five lowest-energy eigenstates with equal weights. At the end of the \dmrgscf computations, the active orbitals were relocalized using the Pipek-Mezey procedure,\cite{Pipek:1989p4916} and the core and external orbitals were canonicalized as the eigenvalues of the state-specific generalized Fock operator (\cref{eq:f_gen}). To compute the \tmpsnevpt and \scmpsnevpt energies, reference MPS for each eigenstate were first computed with $M'$ = 500 and then compressed down to $M$ = 250. For C$_n$H$_{n+2}$ with $n$ = 4, 8, and 16, the remaining error in the reference and correlation energies was estimated to be less than 0.1 \meh. To achieve a similar accuracy for \ce{C24H26}, we used $M$ = 500 in \scmpsnevpt and to compute the $E_{act}^{[+1']}$ and $E_{act}^{[-1']}$ contributions in \tmpsnevpt.

Finally, to assign the spatial symmetries of the excited states, we computed transition dipole moments between the states. Forbidden transitions were assigned to have $A_g$ symmetry, while the allowed transitions were labeled as $B_u$. The $B_u$ states corresponding to a large transition dipole moment were assigned as $B_u^+$, indicating that the electronic excitation involves a change of particle-hole character, whereas the remaining $B_u$ states were labeled as $B_u^-$. For polyenes with $n$ = 4, 8, 16, and 24, the \bup state was found to be the 3$^{rd}$, 5$^{th}$, 7$^{th}$, and 9$^{th}$ singlet eigenstate of the zeroth-order Hamiltonian, respectively. Thus, to compute the \bup excitation energies, we increased the number of states evaluated in the state-averaged DMRG computation with $M'$ = 500 to 9 and 11 for $n$ = 16 and 24, respectively.

\subsubsection{Discussion}

\begin{figure}[t!]
   \includegraphics[width=0.45\textwidth]{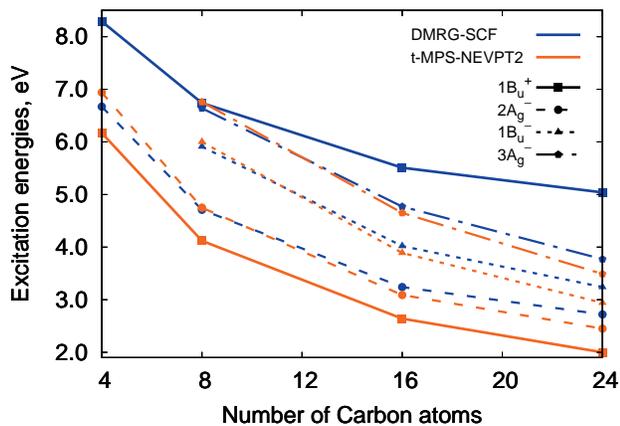}
   \captionsetup{justification=raggedright,singlelinecheck=false}
   \caption{Excitation energies in eV for the \bup, \ag{2}, \bum, and \ag{3} singlet states in the {\it all-trans} conjugated polyenes C$_n$H$_{n+2}$ computed using \dmrgscf and \tmpsnevpt with the cc-pVDZ basis set. Computations employed ($n$e,$n$o) active spaces, where the number of active electrons and orbitals $n$ is equal to the number of C atoms.}
   \label{fig:polyene_spectra}
\end{figure}

\cref{tab:polyene_dz} presents energies, symmetries, and oscillator strengths for the low-lying singlet states of polyenes C$_n$H$_{n+2}$ ($n$ = 4, 8, 16, and 24) computed using the \dmrgscf, \scmpsnevpt, and \tmpsnevpt methods with the cc-pVDZ basis set. In addition, \cref{fig:polyene_spectra} plots the relative energies of the excited states obtained from \dmrgscf and \tmpsnevpt as a function of the chain length $n$. For polyenes with $n$ = 8, 16, and 24, our \dmrgscf computations predict the energies and ordering of the covalent excited states \ag{2} $<$ \bum $<$ \ag{3} in close agreement with the results obtained by Ghosh {\it et al.}\cite{Ghosh:2008p144117} In particular, for all polyenes (including \ce{C4H6}), \dmrgscf predicts the \ag{2} state to be the lowest-energy singlet excited state, while the energy of the ionic \bup state dominated by the HOMO $\rightarrow$ LUMO transition is computed to be 1.6 eV to 2.3 eV higher. 

\begin{table*}[t!]
\begin{flushleft}
\captionsetup{justification=raggedright,singlelinecheck=false}
\caption{Energies, symmetries, and oscillator strengths for the low-lying singlet states in the {\it all-trans} conjugated polyenes C$_n$H$_{n+2}$. Entries for the \ag{1} ground states give the total energies in \eh computed using \dmrgscf, \scmpsnevpt, and \tmpsnevpt with the cc-pVDZ basis set. Entries for the excited states give vertical excitation energies from the ground state in eV. The notation ($n$e,$m$o) denotes the active spaces used in the DMRG and NEVPT2 computations. Oscillator strengths in a.u.\@ were computed using \dmrgscf. The reference numbers in brackets are experimental measurements for substituted polyenes.}
\label{tab:polyene_dz}
\begin{threeparttable}
{
\setstretch{1.2}
\begin{tabular}{C{1.75cm} C{1.5cm} C{2.5cm} C{2.5cm} C{2.5cm} C{2.5cm} C{2.5cm}}
\hline
\hline
Polyene & State & \dmrgscf & \scmpsnevpt & \tmpsnevpt & Oscillator strength & Reference   \\
\hline
\ce{C4H6} 	& \ag{1} 	& $-$154.9772	& $-$155.4862	& $-$155.4866	&  			&    \\
(4e,4o) 		& \bup 	&  8.29		& 6.27 		& 6.17	 	& 2.251 		& 5.92\tnote{a}, 6.21\tnote{b}   \\
			& \ag{2} 	&  6.67 		& 6.96 		& 6.94 		& Forbidden 	& 6.41\tnote{b} \\
\ce{C8H10} 	& \ag{1} 	& $-$308.8259 	& $-$309.8154 	& $-$309.8168 	&  			&    \\
(8e,8o) 		& \bup 	&  6.74		& 4.24		& 4.12		& 3.345 		& 4.41\tnote{c}, 4.76\tnote{d}   \\
			& \ag{2} 	&  4.71		& 4.77 		& 4.75 		& Forbidden 	& 3.59\tnote{e}, 4.81\tnote{d}   \\
			& \bum 	&  5.91		& 6.03 		& 6.00 		& 0.056 		& 5.96\tnote{d}   \\
			& \ag{3} 	&  6.64		& 6.80 		& 6.76 		& Forbidden 	&    \\					
\ce{C16H18} 	& \ag{1} 	& $-$616.5142 	& $-$618.4719 	& $-$618.4754	&  			&    \\
(16e,16o)  	& \bup 	&  5.51		& 2.82 		& 2.64 		& 4.964 		& (2.82)\tnote{f}   \\
			& \ag{2} 	&  3.24		& 3.14 		& 3.09	 	& Forbidden 	& (2.21)\tnote{f}   \\
			& \bum 	&  4.02		& 3.95		& 3.89		& 0.031 		&    \\
			& \ag{3} 	&  4.77		& 4.73 		& 4.65 		& Forbidden 	&    \\
\ce{C24H26} 	& \ag{1} 	& $-$924.2014 	& $-$927.1284 	& $-$927.1354	&  			&    \\
(24e,24o)  	& \bup 	&  5.04		& 2.25 		& 2.00 		& 6.167 		& (2.25)\tnote{g}   \\
			& \ag{2} 	&  2.72		& 2.54 		& 2.45 		& Forbidden 	& (1.51)\tnote{g}   \\
			& \bum 	&  3.24		& 3.08 		& 2.94 		& 0.025 		& (1.80)\tnote{g}   \\
			& \ag{3} 	&  3.77		& 3.63 		& 3.49 		& Forbidden 	& (2.04)\tnote{g}   \\
\hline
\hline
\end{tabular}
\begin{tablenotes}
\item[a] Experimental adiabatic excitation energy.\cite{Doering:1981p2477,Doering:1980p3617}
\item[b] Best theoretical vertical excitation energy.\cite{Watson:2012p4013}
\item[c] Experimental adiabatic excitation energy.\cite{Heimbrook:1981p4338,Heimbrook:1984p1592,Leopold:1984p4210}
\item[d] Best theoretical vertical excitation energy.\cite{Angeli:2011p184302}
\item[e] Experimental adiabatic excitation energy.\cite{Petek:1993p3777}
\item[f] Experimental adiabatic excitation energy for a substituted polyene.\cite{Kohler:1988p5422}
\item[g] Experimental adiabatic excitation energy for a substituted polyene.\cite{Furuichi:2002p547}
\end{tablenotes}
}
\end{threeparttable}
\end{flushleft}
\end{table*}

Including the dynamic correlation lowers the relative energy of the ionic \bup state drastically. This is seen in the difference of the \bup excitation energies computed using \scmpsnevpt and \dmrgscf, which ranges from 2 eV in \ce{C4H6} to 2.7 eV in \ce{C24H26}, reducing the excitation energy for the longer polyene by more than a factor of two. Similar results are obtained using the uncontracted \tmpsnevpt method, which lowers the \bup energy by an additional 0.10 to 0.25 eV relative to \scmpsnevpt, due to the removal of the strong contraction approximation. Both methods predict \bup to be the lowest-energy singlet excited state. The effect of dynamic correlation on the relative energies of the covalent states \ag{2}, \bum, and \ag{3} is much smaller (\cref{fig:polyene_spectra}). Nevertheless, the correlation contributions to the relative energies of these states increase with chain length and become significant for longer polyenes, reaching $\sim$ 0.3 eV in \ce{C24H26} at the \tmpsnevpt level of theory. Interestingly, in shorter polyenes (\ce{C4H6} and \ce{C8H10}), including the dynamic correlation actually leads to an increase of the excitation energies for the covalent states computed using \scmpsnevpt and \tmpsnevpt relative to \dmrgscf, indicating that the correlation is more important in the ground rather than the excited electronic states for these molecules. As we describe below, a similar trend is observed even when the larger aug-cc-pVTZ basis set is used in the computations. For longer polyenes, dynamic correlation lowers the vertical excitation energies for the covalent transitions. As a result, the \scmpsnevpt and \tmpsnevpt excitation energies decrease with increasing chain length $n$ faster than compared to \dmrgscf, as shown in \cref{fig:polyene_spectra}. 

\begin{figure}[t!]
   \includegraphics[width=0.45\textwidth]{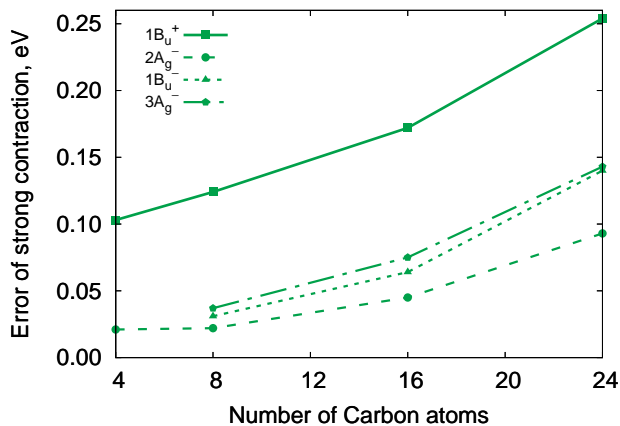}
   \captionsetup{justification=raggedright,singlelinecheck=false}
   \caption{Strong contraction errors in excitation energies (eV) for the lowest-lying singlet states of the {\it all-trans} conjugated polyenes C$_n$H$_{n+2}$ computed as the difference between the \scmpsnevpt and \tmpsnevpt results.}
   \label{fig:polyene_contraction}
\end{figure}

\cref{fig:polyene_contraction} plots the error of the strong contraction approximation in the excitation energies of polyenes with the number of carbon atoms computed as the difference between the \scmpsnevpt and \tmpsnevpt results. Although in the shorter polyenes (\ce{C4H6} and \ce{C8H10}) strong contraction mainly affects the energy of the ionic \bup state, in the longer polyenes the errors of strong contraction become significant even for the covalent \ag{2}, \bum, and \ag{3} states. In particular, for \ce{C24H26}, the strong contraction errors amount to 0.09 to 0.14 eV in the excitation energy for the covalent transitions, which is comparable to the $\sim$ 0.15 eV lowering of the energies due to dynamic correlation computed by \scmpsnevpt.

\begin{table*}[t!]
\begin{flushleft}
\captionsetup{justification=raggedright,singlelinecheck=false}
\caption{Energies for the lowest-lying singlet states in the {\it all-trans} conjugated polyenes \ce{C4H6} and \ce{C8H10}. Entries for the \ag{1} ground states give the total energies in \eh computed using CASSCF, sc-NEVPT2, and t-NEVPT2 with the aug-cc-pVTZ basis set. Entries for the excited states give vertical excitation energies from the ground state in eV. The notation ($n$e,$m$o) denotes the active spaces used in the CASSCF computations.}
\label{tab:polyene_atz}
\begin{threeparttable}
{
\setstretch{1.2}
\begin{tabular}{C{2cm} C{2cm} C{3cm} C{3cm} C{3cm} C{3cm}}
\hline
\hline
Polyene & State & CASSCF & sc-NEVPT2 & t-NEVPT2 & Reference \\
\hline
\ce{C4H6} 	& \ag{1} 	& $-$154.9845	& $-$155.7199	& $-$155.7211	&   \\
(4e,4o)		& \bup 	& 7.33 		& 6.06 		& 5.83 		&5.92\tnote{a}, 6.21\tnote{b}\\
 			& \ag{2} 	& 5.40 		& 6.61 		& 6.62 		&6.41\tnote{b}\\
\ce{C8H10} 	& \ag{1} 	&  $-$308.9070	& $-$310.2677 	& $-$310.2694 	&\\
(8e,8o) 		& \bup 	& 6.65		& 4.02		& 3.79		&4.41\tnote{c}, 4.76\tnote{d}\\
			& \ag{2} 	& 4.79 		& 4.81 		& 4.78 		&3.59\tnote{e}, 4.81\tnote{d}\\
			& \bum 	& 6.00 		& 6.07 		& 6.02 		&5.96\tnote{d}\\
			& \ag{3} 	& 6.74 		& 6.82 		& 6.75 		&\\					
\hline
\hline
\end{tabular}
\begin{tablenotes}
\item[a] Experimental adiabatic excitation energy.\cite{Doering:1981p2477,Doering:1980p3617}
\item[b] Best theoretical vertical excitation energy.\cite{Watson:2012p4013}
\item[c] Experimental adiabatic excitation energy.\cite{Heimbrook:1981p4338,Heimbrook:1984p1592,Leopold:1984p4210}
\item[d] Best theoretical vertical excitation energy.\cite{Angeli:2011p184302}
\item[e] Experimental adiabatic excitation energy.\cite{Petek:1993p3777}
\end{tablenotes}
}
\end{threeparttable}
\end{flushleft}
\end{table*}

To assess the relative performance of \scmpsnevpt and \tmpsnevpt for the description of excited states with ionic and covalent character, we computed the \ce{C4H6} and \ce{C8H10} vertical excitation energies using sc-NEVPT2 and t-NEVPT2 with the large aug-cc-pVTZ basis set and CASSCF reference wavefunctions (\cref{tab:polyene_atz}). Increasing the basis set from cc-pVDZ to aug-cc-pVTZ lowers the t-NEVPT2 relative energies of the \bup and \ag{2} states in \ce{C4H6} by 0.34 and 0.32 eV, respectively. For the covalent \ag{2} state, both sc-NEVPT2 and t-NEVPT2 predict excitation energies of $\sim$ 6.6 eV, in a reasonable agreement with a reference value of 6.41 eV from highly accurate coupled cluster computations.\cite{Watson:2012p4013} Considering the strong basis set dependence of the \ag{2} excitation energy, we expect that this agreement will improve at the CBS limit. On the other hand, the sc-NEVPT2 and t-NEVPT2 vertical excitation energies for the \bup state (6.06 and 5.83 eV) are too low relative to the best available theoretical value of 6.21 eV (\cref{tab:polyene_atz}), indicating that the two methods will significantly underestimate the excitation energy of the ionic state at the CBS limit. A similar performance of sc-NEVPT2 and t-NEVPT2 is observed for \ce{C8H10}. In this case, both methods yield excitation energies for the covalent states \ag{2} and \bum that are in the excellent agreement with the best available theoretical reference values,\cite{Angeli:2011p184302} while the relative energy of the ionic \bup state is underestimated by more than 0.7 eV. Similar errors for the \bup state have been observed by Angeli and Pastore in the study of \ce{C8H10} using the second-order quasi-degenerate perturbation theory (QD-PT2) with the (8e,8o) active space.\cite{Angeli:2011p184302} They demonstrated that increasing the active space from (8e,8o) to (8e,16o) results in $\sim$ 0.7 eV increase in the \bup excitation energy computed using QD-PT2, leading to a closer agreement with the experiment. 

The underestimation of the \bup relative energies due to insufficiently large active spaces may explain the dependence of the excitation energies with the chain length $n$ computed using \tmpsnevpt (\cref{fig:polyene_spectra}), where we observe no crossing of \bup with the covalent \ag{2} and \bum transitions, contrary to experimental results on substituted polyenes\cite{Furuichi:2002p547,Kohler:1988p5422} and some of calculations.\cite{Kurashige:2004p425,Schmidt:2012p124309} Re-investigating the trends in the excitation energies of the long polyenes with a double $\pi$-active space, as employed by Angeli and Pastore, will be the subject of future work.

\section{Conclusions}
\label{sec:conclusions}

In this work, we developed an efficient algorithm to describe dynamic correlation in strongly correlated systems with large active spaces and basis sets based on a combination of the time-dependent second-order $N$-electron valence perturbation theory (t-NEVPT2) with matrix product state (MPS) reference wavefunctions. The resulting \tmpsnevpt approach is equivalent to the fully uncontracted $N$-electron valence perturbation theory, but can efficiently compute correlation energies using the time-dependent density matrix renormalization group algorithm. In addition, we presented a new MPS-based implementation of strongly-contracted NEVPT2 (\scmpsnevpt) that has a lower computational scaling than the commonly used internally-contracted NEVPT2 variants. Using these new methods, we computed the dissociation energy of the chromium dimer and investigated the importance of dynamic correlation in the low-lying excited states of {\it all-trans} polyenes (\ce{C4H6} to \ce{C24H26}). For the chromium dimer, the active space used included the ``double d'' shell of the chromium atoms. Our \ce{Cr2} dissociation energy computed using the uncontracted \tmpsnevpt method is in a good agreement with the experimental data and the previously reported results from strongly-contracted NEVPT2.\cite{Guo:2016p1583} In a study of polyenes, using a complete $\pi$-valence active space, our results demonstrate that including dynamic correlation is very important for the ionic $\pi$-$\pi^*$ transitions, while it is less important for excitations with covalent character. In particular, for the \ce{C24H26} polyene, incorporating dynamic correlation at the \tmpsnevpt level of theory lowers the energy of the ionic transition by $\sim$ 3 eV, whereas only $\sim$ 0.3 eV change is observed for the covalent electronic states. Our results suggest that both \scmpsnevpt and \tmpsnevpt significantly underestimate the relative energy of the ionic state when combined with the complete $\pi$-valence active space. In this case, using the ``double'' $\pi$-active space is expected to improve the description of the excitation energies and will be the subject of our future work. Overall, our results demonstrate that \tmpsnevpt and \scmpsnevpt are promising methods for the study of strongly correlated systems that can be applied to a variety of challenging problems.

\section{Acknowledgements}
This work was supported by the U.S.\@ Department of Energy (DOE), Office of Science through
Award DE-SC0008624 (primary support for A.Y.S.), and Award DE-SC0010530 (additional support for G.K.-L.C.), and  used resources of the National Energy Research Scientific Computing Center, a DOE Office of Science User Facility supported by the Office of Science of the U.S. Department of Energy under Contract No.\@ DE-AC02-05CH11231. The authors would like to thank Dr.\@ Zhendong Li for insightful discussions.

\section{Appendix A: Spin-orbital two-body Green's functions in spin-adapted DMRG}
\label{sec:appendix}
Here, we describe how to compute the spin-orbital 2-GF using the spin-adapted DMRG algorithm. 
First, we evaluate the elements of $\mathbf{G}\boldsymbol{^{[-2]}}(\tau)$, which in the spin-orbital basis have the following form: $G_{z_{\kappa}w_{\lambda}}^{x_{\rho}y_{\sigma}}(\tau) = \braket{\Psi_0| \c{x_{\rho}} (\tau) \c{y_{\sigma}} (\tau) \a{w_{\lambda}} \a{z_{\kappa}} |\Psi_0} \equiv \braket{\c{x\rho} (\tau) \c{y\sigma} (\tau) \a{w\lambda} \a{z\kappa}}$, where $x,y,w,z$ denote spatial orbitals and $\rho,\sigma,\kappa,\lambda$ denote the spin of these orbitals (e.g., $\rho=\alpha,\beta$). As discussed in \cref{sec:t_mps_nevpt2_spin_adaptation}, in spin-adapted DMRG $\c{x}$ and $\a{x}$ are spin tensor creation and annihilation operators. We use a shorthand notation for the product of two tensor operators $\hat{A}_{xy}^{S,M}\equiv[\c{x} \c{y} ]^{S}_{M}$, where square brackets denote spin coupling ($S=0,1$). The adjoint of a spin tensor operator is defined with an additional sign factor to preserve the Condon-Shortley phase convention: $\hat{A}_{xy}^{S,M\ddagger}=(-1)^{S+M}\hat{A}_{xy}^{S,-M\dagger}$. In addition, we define the spin-adapted tensor product $\otimes_S$ as $\hat{A}_{1}^{S_1}\otimes_S\hat{A}_{2}^{S_2}=[\hat{A}_{1}^{S_1}\hat{A}_{2}^{S_2}]^S$, where $[\hat{A}_{1}^{S_1}\hat{A}_{2}^{S_2}]^S_M = \sum_{M_1M_2}c_{M_1,M_2,M}^{S_1,S_2,S}(\hat{A}_{1})^{S_1}_{M_1}(\hat{A}_{2})^{S_2}_{M_2}$ and $c_{M_1,M_2,M}^{S_1,S_2,S}$ are Clebsch-Gordan coefficients. Thus, the elements $G_{z_{\kappa}w_{\lambda}}^{x_{\rho}y_{\sigma}}(\tau)$ can be computed by solving the linear system of equations:
\begin{widetext}
\begin{align}
	\label{eq:gf_ccaa_sa_to_so}
	\begin{pmatrix} 
	0 & \frac{1}{2} & \frac{1}{2} & -\frac{1}{2} & -\frac{1}{2} & 0 \\
	\frac{1}{\sqrt{3}} & \frac{1}{\sqrt{12}} & \frac{1}{\sqrt{12}} & \frac{1}{\sqrt{12}} & \frac{1}{\sqrt{12}} & \frac{1}{\sqrt{3}}  \\
 	0 & -\frac{1}{2} & \frac{1}{2} & \frac{1}{2} & -\frac{1}{2} & 0 \\
 	0 & \frac{1}{2} & -\frac{1}{2} & \frac{1}{2} & -\frac{1}{2} & 0 \\
 	\frac{1}{\sqrt{2}} & 0 & 0 & 0 & 0 & -\frac{1}{\sqrt{2}} \\
 	\frac{1}{\sqrt{6}} & -\frac{1}{\sqrt{6}} & -\frac{1}{\sqrt{6}} & -\frac{1}{\sqrt{6}} & -\frac{1}{\sqrt{6}} & \frac{1}{\sqrt{6}} 
	\end{pmatrix}
	\begin{pmatrix} 
	 \braket{\c{x\alpha} (\tau) \c{y\alpha} (\tau) \a{w\alpha} \a{z\alpha}} \\
	 \braket{\c{x\alpha} (\tau) \c{y\beta} (\tau) \a{w\beta} \a{z\alpha}} \\
	 \braket{\c{x\beta} (\tau) \c{y\alpha} (\tau) \a{w\alpha} \a{z\beta}} \\
	 \braket{\c{x\beta} (\tau) \c{y\alpha} (\tau) \a{w\beta} \a{z\alpha}} \\
	 \braket{\c{x\alpha} (\tau) \c{y\beta} (\tau) \a{w\alpha} \a{z\beta}} \\
	 \braket{\c{x\beta} (\tau) \c{y\beta} (\tau) \a{w\beta} \a{z\beta}} 
	\end{pmatrix}
	=
	\begin{pmatrix} 
	 \braket{\hat{A}_{xy}^{0} (\tau) \otimes_0 \hat{A}_{zw}^{0\ddagger}} \\
	 \braket{\hat{A}_{xy}^{1} (\tau) \otimes_0 \hat{A}_{zw}^{1\ddagger}} \\
	 \braket{\hat{A}_{xy}^{0} (\tau) \otimes_1 \hat{A}_{zw}^{1\ddagger}} \\
	 \braket{\hat{A}_{xy}^{1} (\tau) \otimes_1 \hat{A}_{zw}^{0\ddagger}} \\
	 \braket{\hat{A}_{xy}^{1} (\tau) \otimes_1 \hat{A}_{zw}^{1\ddagger}} \\
	 \braket{\hat{A}_{xy}^{1} (\tau) \otimes_2 \hat{A}_{zw}^{1\ddagger}} 
	\end{pmatrix}
	 \ .
\end{align}
\cref{eq:gf_ccaa_sa_to_so} can also be used to compute elements of $\mathbf{G}\boldsymbol{^{[+2]}}(\tau)$ by replacing creation operators with annihilation operators and vice versa, e.g. $\braket{\c{x\alpha} (\tau) \c{y\beta} (\tau) \a{w\beta} \a{z\alpha}}\rightarrow\braket{\a{x\alpha} (\tau) \a{y\beta} (\tau) \c{w\beta} \c{z\alpha}}$ and $\braket{\hat{A}_{xy}^{1} (\tau) \otimes_0 \hat{A}_{zw}^{1\ddagger}}\rightarrow\braket{\hat{A}_{yx}^{1\ddagger} (\tau) \otimes_0 \hat{A}_{wz}^{1}}$. To compute elements of $\mathbf{G}\boldsymbol{^{[0']}}(\tau)$, we define a product of tensor operators $\hat{B}_{xy}^{S,M}\equiv[\c{x} \a{y} ]^{S}_{M}$. In this case, the system of linear equations has the form:
\begin{align}
	\label{eq:gf_caca_sa_to_so}
	\begin{pmatrix} 
	\frac{1}{2} & \frac{1}{2} & 0 & 0 & \frac{1}{2} & \frac{1}{2} \\
	\frac{1}{\sqrt{12}} & -\frac{1}{\sqrt{12}} & \frac{1}{\sqrt{3}} & \frac{1}{\sqrt{3}} & -\frac{1}{\sqrt{12}} & \frac{1}{\sqrt{12}} \\
 	-\frac{1}{2} & \frac{1}{2} & 0 & 0 & -\frac{1}{2} & \frac{1}{2} \\
 	\frac{1}{2} & \frac{1}{2} & 0 & 0 & -\frac{1}{2} & -\frac{1}{2} \\
 	0 & 0 & \frac{1}{\sqrt{2}} & -\frac{1}{\sqrt{2}} & 0 & 0 \\
 	-\frac{1}{\sqrt{6}} & \frac{1}{\sqrt{6}} & \frac{1}{\sqrt{6}} & \frac{1}{\sqrt{6}} & \frac{1}{\sqrt{6}} & -\frac{1}{\sqrt{6}}
	\end{pmatrix}
	\begin{pmatrix} 
	 \braket{\c{x\alpha} (\tau) \a{y\alpha} (\tau) \c{w\alpha} \a{z\alpha}} \\
	 \braket{\c{x\alpha} (\tau) \a{y\alpha} (\tau) \c{w\beta} \a{z\beta}} \\
	 \braket{\c{x\alpha} (\tau) \a{y\beta} (\tau) \c{w\beta} \a{z\alpha}} \\
	 \braket{\c{x\beta} (\tau) \a{y\alpha} (\tau) \c{w\alpha} \a{z\beta}} \\
	 \braket{\c{x\beta} (\tau) \a{y\beta} (\tau) \c{w\alpha} \a{z\alpha}} \\
	 \braket{\c{x\beta} (\tau) \a{y\beta} (\tau) \c{w\beta} \a{z\beta}} 
	\end{pmatrix}
	=
	\begin{pmatrix} 
	 \braket{\hat{B}_{xy}^{0} (\tau) \otimes_0 \hat{B}_{zw}^{0\ddagger}} \\
	 \braket{\hat{B}_{xy}^{1} (\tau) \otimes_0 \hat{B}_{zw}^{1\ddagger}} \\
	 \braket{\hat{B}_{xy}^{0} (\tau) \otimes_1 \hat{B}_{zw}^{1\ddagger}} \\
	 \braket{\hat{B}_{xy}^{1} (\tau) \otimes_1 \hat{B}_{zw}^{0\ddagger}} \\
	 \braket{\hat{B}_{xy}^{1} (\tau) \otimes_1 \hat{B}_{zw}^{1\ddagger}} \\
	 \braket{\hat{B}_{xy}^{1} (\tau) \otimes_2 \hat{B}_{zw}^{1\ddagger}} 
	\end{pmatrix}
	 \ .
\end{align}

\cref{eq:gf_ccaa_sa_to_so,eq:gf_caca_sa_to_so} can be simplified for a closed-shell reference wavefunction $\ket{\Psi_0}$ to obtain compact expressions:
\begin{align}
	\label{eq:gf_sa_to_so_simple_1}
	\braket{\c{x\alpha} (\tau) \c{y\alpha} (\tau) \a{w\alpha} \a{z\alpha}} 
	&= \frac{1}{\sqrt{3}}\braket{\hat{A}_{xy}^{1} (\tau) \otimes_0 \hat{A}_{zw}^{1\ddagger}} \ , \\
	\label{eq:gf_sa_to_so_simple_2}
	\braket{\c{x\alpha} (\tau) \c{y\beta} (\tau) \a{w\beta} \a{z\alpha}}
	&= \frac{1}{\sqrt{12}} \braket{\hat{A}_{xy}^{1} (\tau) \otimes_0 \hat{A}_{zw}^{1\ddagger}} + \frac{1}{2}\braket{\hat{A}_{xy}^{0} (\tau) \otimes_0 \hat{A}_{zw}^{0\ddagger}} \ , \\
	\label{eq:gf_sa_to_so_simple_3}
	\braket{\c{x\beta} (\tau) \c{y\alpha} (\tau) \a{w\beta} \a{z\alpha}}
	&= \frac{1}{\sqrt{12}} \braket{\hat{A}_{xy}^{1} (\tau) \otimes_0 \hat{A}_{zw}^{1\ddagger}} - \frac{1}{2}\braket{\hat{A}_{xy}^{0} (\tau) \otimes_0 \hat{A}_{zw}^{0\ddagger}} \ , \\
	\label{eq:gf_sa_to_so_simple_4}
	 \braket{\c{x\alpha} (\tau) \a{y\beta} (\tau) \c{w\beta} \a{z\alpha}} 
	 &= \frac{1}{\sqrt{3}} \braket{\hat{B}_{xy}^{1} (\tau) \otimes_0 \hat{B}_{zw}^{1\ddagger}} \ , \\
	\label{eq:gf_sa_to_so_simple_5}
	 \braket{\c{x\alpha} (\tau) \a{y\alpha} (\tau) \c{w\alpha} \a{z\alpha}} 
	 &= \frac{1}{2}\braket{\hat{B}_{xy}^{0} (\tau) \otimes_0 \hat{B}_{zw}^{0\ddagger}} + \frac{1}{\sqrt{12}} \braket{\hat{B}_{xy}^{1} (\tau) \otimes_0 \hat{B}_{zw}^{1\ddagger}} \ , \\
	\label{eq:gf_sa_to_so_simple_6}
	 \braket{\c{x\alpha} (\tau) \a{y\alpha} (\tau) \c{w\beta} \a{z\beta}} 
	 &= \frac{1}{2}\braket{\hat{B}_{xy}^{0} (\tau) \otimes_0 \hat{B}_{zw}^{0\ddagger}} - \frac{1}{\sqrt{12}} \braket{\hat{B}_{xy}^{1} (\tau) \otimes_0 \hat{B}_{zw}^{1\ddagger}} \ .
\end{align}
The remaining spin-orbital elements can be obtained using the following symmetry relations: $\braket{\c{x\alpha} (\tau) \c{y\alpha} (\tau) \a{w\alpha} \a{z\alpha}} = \braket{\c{x\beta} (\tau) \c{y\beta} (\tau) \a{w\beta} \a{z\beta}}$, $\braket{\c{x\alpha} (\tau) \c{y\beta} (\tau) \a{w\beta} \a{z\alpha}} = \braket{\c{x\beta} (\tau) \c{y\alpha} (\tau) \a{w\alpha} \a{z\beta}}$, $\braket{\c{x\beta} (\tau) \c{y\alpha} (\tau) \a{w\beta} \a{z\alpha}} = \braket{\c{x\alpha} (\tau) \c{y\beta} (\tau) \a{w\alpha} \a{z\beta}}$, $\braket{\c{x\alpha} (\tau) \a{y\alpha} (\tau) \c{w\alpha} \a{z\alpha}} = \braket{\c{x\beta} (\tau) \a{y\beta} (\tau) \c{w\beta} \a{z\beta}}$, $\braket{\c{x\alpha} (\tau) \a{y\beta} (\tau) \c{w\beta} \a{z\alpha}} = \braket{\c{x\beta} (\tau) \a{y\alpha} (\tau) \c{w\alpha} \a{z\beta}}$, $\braket{\c{x\alpha} (\tau) \a{y\alpha} (\tau) \c{w\beta} \a{z\beta}} = \braket{\c{x\beta} (\tau) \a{y\beta} (\tau) \c{w\alpha} \a{z\alpha}}$.
\end{widetext}

\section{Appendix B: Strongly-contracted NEVPT2 with MPS compression ($\mbox{\scmpsnevpt}$)}
\label{sec:sc_nevpt2}

In this section, we describe an efficient implementation of strongly-contracted NEVPT2 with MPS reference wavefunctions (\scmpsnevpt). Although a sc-NEVPT2 implementation on top of MPS wavefunctions was reported in Ref.~\citenum{Guo:2016p1583},  here we additionally introduce MPS compression, which greatly reduces the computational cost, and allows for larger active spaces to be treated. We will not describe the general theory of sc-NEVPT2 in detail, but instead refer the reader to Refs.~\citenum{Angeli:2001p10252} and \citenum{Angeli:2001p297}.

In sc-NEVPT2, for each of the 7 subspaces ($i\in\{+1$, $-1$, $+2$, $-2$, $+1'$, $-1'$, $0'\}$) described in \cref{sec:t_nevpt2},
the first-order wavefunction is expanded in a basis consisting
of a single perturber function for each core or external index.
The perturber function is constructed by fixing  (a set of)
of core or external indices in $\hat{V}$ and acting on $\ket{\Psi_0}$. For example, 
in the  $[+1']$ subclass, the perturber functions are $\ket{\Psi_i^{[+1']}} = \hat{h}^{\dag}_{i} a_i^{}\ket{\Psi_0}$. Because the perturber functions are
all orthogonal to each other, they lead to very simple expressions
for the second-order energy.
For example, the perturbation contribution in the $[+1']$ subclass is
\begin{align}
	\label{eq:sc_e+1}
	E^{[+1']} 
  &=  \sum_{i}  \frac{|\braket{\Psi_0| a^\dag_i \hat{h}_{i}|\tilde{\Psi}^{[+1']}_i}|^2}{E_{\tilde{\Psi}_i^{[+1']}} -E_0}  \notag \\
  &=  \sum_{i}  \frac{||\ket{{\Psi}_i^{[+1']}}||^2}{E_{\tilde{\Psi}_i^{[+1']}} -E_0} \ ,
\end{align}
where $\ket{\tilde{\Psi}_i^{[+1']}} = \frac{\ket{\Psi_i^{[+1']}}}{||\ket{\Psi_i^{[+1']}}||}$, $E_{\tilde{\Psi}_i^{[+1']}} = \frac{\braket{\Psi_i^{[+1']}| \hat{H}_D|\Psi_i^{[+1']}}}{\braket{\Psi_i^{[+1']}|\Psi_i^{[+1']}}}$.

The main bottleneck in obtaining the sc-NEVPT2 energy is computing the term $\braket{\Psi_i^{[+1']}| \hat{H}_D|\Psi_i^{[+1']}}$ and
the analogous contribution for $\ket{\Psi_a^{[-1']}}$. This corresponds to
\begin{align}
  \braket{\Psi_i^{[+1']}| \hat{H}_D|\Psi_i^{[+1']}} &= \braket{\Psi_0|a^{\dag}_i\hat{h}^{}_{i}\hat{H}_D\hat{h}^{\dag}_{i}a_i|\Psi_0} \notag \\
  &= \braket{\Psi_0|\hat{h}^{}_{i}\hat{H}_D\hat{h}^{\dag}_{i}|\Psi_0} - \epsilon_i\braket{\Psi_0|\hat{h}^{}_{i}\hat{h}^{\dag}_{i}|\Psi_0} \notag \\
  &= \braket{\Psi_0|\hat{h}^{}_{i}[\hat{H}_D,\hat{h}^{\dag}_{i}]|\Psi_0} \notag \\
  &+ (E_0 -\epsilon_i)\braket{\Psi_0|\hat{h}^{}_{i}\hat{h}^{\dag}_{i}|\Psi_0} \ .
\end{align}
The above object involves 8 active indices, and the expectation value can be computed from the contraction of the 4-RDM and two-electron integrals.~\cite{Angeli:2001p10252, Angeli:2001p297, Angeli:2002ik}

However, similarly to how we avoid calculating the 3-GF in \cref{eq:e_+1p,eq:e_-1p} in t-NEVPT2, we can avoid calculating the 4-RDM in sc-NEVPT2. 
We can represent $\ket{\Psi_i^{[+1']}}$ and $\ket{\Psi_a^{[-1']}}$ as MPS of bond dimension $M_1$, by carrying out a variational compression as
described in \cref{sec:t_mps_nevpt2_gf}
to minimize  $|| \ket{\Psi_i^{[+1']}} - \hat{h}_i^\dag a_i^{}\ket{\Psi_0}||$ and $|| \ket{\Psi_a^{[-1']}} - a_a^{\dag} \hat{h}^{}_a \ket{\Psi_0}||$.
Assuming $M_1 \sim M_0$, the cost of carrying out compressions for all the perturbers in class $[+1']$ and $[-1']$ is 
$\mathcal{O}(N_{ext}N^2_{act}M_0^3)$. The computational cost for the subsequent $\hat{H}_D$ expectation value is  $\mathcal{O}(N_{ext}N^3_{act}M_0^3)$, however
as no sweeps are required, this has a very low prefactor and in our calculations the amount of time in this step is negligible. Thus, the
computational cost of sc-NEVPT2 with MPS compression (\scmpsnevpt) scales in practice as $\mathcal{O}(N_{act}^4M_0^3+N_{act}^6M_0^2+N_{ext}N^2_{act}M_0^3)$. This is significantly lower
than the $\mathcal{O}(N_{act}^5M_0^3+N_{act}^8M_0^2)$ cost of computing the 4-RDM, which dominates the standard sc-NEVPT2 algorithm presented in Ref.~\citenum{Guo:2016p1583}.


\end{document}